\documentclass[pre,aps,onecolumn,superscriptaddress,notitlepage,nofootinbib]{revtex4-1}
\usepackage{amssymb,epsfig,amsmath,bm,subfigure,graphicx,textcomp,url,float,color,cases}

\usepackage{epsfig}
\usepackage{textcomp}
\usepackage[normalem]{ulem}

\usepackage[colorlinks=true, urlcolor=blue, anchorcolor=blue, citecolor=blue,filecolor=blue,linkcolor=blue,menucolor=blue
]{hyperref}

\usepackage{color}

\usepackage[titletoc,title]{appendix}

\begin{document}

\title{Large deviations and phase transitions in spectral linear statistics
of Gaussian random matrices}
\author{Alexander Valov}
\email{aleksandr.valov@mail.huji.ac.il}
\author{Baruch Meerson}
\email{meerson@mail.huji.ac.il}
\affiliation{Racah Institute of Physics, Hebrew University of Jerusalem, Jerusalem 91904, Israel}
\author{Pavel V. Sasorov}
\email{pavel.sasorov@gmail.com}
\affiliation{Institute of Physics CAS, ELI Beamlines, 182 21 Prague, Czech Republic}

\begin{abstract}
We evaluate, in the large-$N$ limit, the complete probability distribution $\mathcal{P}(A,m)$ of the values $A$ of the sum $\sum_{i=1}^{N} |\lambda_i|^m$, where $\lambda_i$ ($i=1,2,\dots, N$) are the eigenvalues of a Gaussian random matrix, and $m$ is a positive real number. Combining the Coulomb gas method with numerical simulations using a matrix variant of the Wang-Landau algorithm, we found that, in the limit of $N\to \infty$, the rate function of $\mathcal{P}(A,m)$ exhibits phase transitions of different characters. The phase diagram of the system on the $(A,m)$ plane is surprisingly rich, as it includes three regions: (i) a region with a single-interval support of the optimal spectrum of eigenvalues, (ii) a region emerging for $m<2$ where the optimal spectrum splits into two separate intervals, and (iii) a region emerging for $m>2$ where the maximum or minimum eigenvalue ``evaporates" from the rest of eigenvalues and dominates the statistics of $A$. The phase transition between regions (i) and (iii) is of second order.
Analytical arguments and numerical simulations strongly suggest that the phase transition between regions (i) and (ii) is of (in general) fractional order
$p=1+1/|m-1|$, where $0<m<2$. The transition becomes of infinite order in the special case of $m=1$, where we provide a more complete analytical and numerical description. Remarkably, the transition between regions (i) and (ii) for $m\leq 1$ and the transition between regions (i) and (iii) for $m>2$ occur at the ground state of the Coulomb gas which corresponds to the Wigner's semicircular distribution.
\end{abstract}
\maketitle
\tableofcontents

\section{Introduction and formalism}
\label{intro}

Understanding statistical properties of sums of large numbers of strongly correlated random variables is a classical
problem in probability theory with important applications in natural sciences and beyond. Important examples
of strongly correlated random variables are provided by the real eigenvalues $\lambda_i$ of the
classical $\beta$-Gaussian, $\beta$-Wishart and $\beta$-Jacobi ensembles of random matrices \cite{Forrester}.
Random matrix theory has numerous applications ranging from nuclear physics to  condensed matter physics, quantum information and quantum chromodynamics, deep learning and finance, to name just a few \cite{Akemann,Potters}. Here a natural question is about
spectral linear statistics, that is the probability distribution of the values $A$ of sums of the type $\sum_i^N f(\lambda_i)$,
where $f(\dots)$ is a given function, and $N\gg 1$. Not surprisingly, spectral linear statistics of large random matrices have attracted much attention in the last 20 years, see Refs.  \cite{Cunden,Cunden2014,Cunden2016,GrabschTexier2016} and references therein. Another important class of systems for which large deviations of linear statistics is of interest, and can be studied by similar methods, is provided by a family of non-interacting and interacting spinless fermions in confining potentials. Their close connection to random matrices is well documented \cite{Dean_2019,Smith_2021}.

In this work we consider $\beta$-Gaussian ensembles of large random matrices and study the probability distribution $\mathcal{P}(A,m)$ of the values $A$ taken by the sum of absolute values of the eigenvalues raised to an arbitrary real power $m>0$:
\begin{equation}\label{exactsum}
\sum_{i=1}^{N} |\lambda_i|^m = A>0\,.
\end{equation}
The linear statistics (\ref{exactsum}) is a particular case of the general linear statistics
\begin{equation}\label{generalsum}
\sum_{i=1}^{N} f(\lambda_i) = A\,.
\end{equation}
A typical behavior of the fluctuating quantity $A$ from Eq.~(\ref{generalsum}) -- as described by the mean $\bar{A}$ and the variance $\text{var}(A)$ -- has been studied in many papers \cite{Dyson,Beenakker1,Beenakker2,Politzer,Chen,Baker,Soshnikov,Lytova,Basor,Pastur,Johansson,Costin,FL} starting from the classical work of Dyson and Mehta \cite{Dyson}. Here we focus on large deviations of $A$, defined by Eq.~(\ref{exactsum}), which correspond to the large-$A$ tails of $\mathcal{P}(A,m)$.  They have received much less attention, except for the special case of $m=2$, where $\mathcal{P}(A,2)$ -- a gamma-distribution -- is known exactly, see \textit{e.g.} Ref. \cite{Grela}\footnote{\label{texierandsatya}A similar in spirit problem appears in the case of a Coulomb gas of $N$ particles confined by the potential $V(x)=x-\ln x$ at $x>0$. Large deviations of the linear statistics $\sum_i^N x_i^{-1}$ of such a gas were studied in Ref. \cite{Texier_2013}.}.

Our approach to this problem is based on a standard Coulomb gas technique, see \textit{e.g.} Ref. \cite{Forrester}. It exploits the fact that the joint probability distribution of the eigenvalues of the (unconstrained) Gaussian $\beta$-ensemble,
\begin{equation}
P_{\beta}(\lambda_1,\lambda_2,\ldots,\lambda_N)\propto\prod\limits_{1\leq i<j<N}\vert \lambda_i-\lambda_j\vert^{\beta}\prod\limits_{i=1}^{N}e^{-\frac{\beta}{2} \lambda_i^2}\,,
\label{jpdf}
\end{equation}
can be interpreted as a measure of  a 2D Coulomb gas of $N$ particles  confined on the line by a one-dimensional harmonic potential:
\begin{equation}
P_{\beta}(\lambda_1,\lambda_2,\ldots,\lambda_N)\propto \exp\left[-\beta E(\bm{\lambda}) \right]\,,
\end{equation}
where $\beta$ plays the role of inverse temperature, and the energy of the Coulomb gas is
\begin{equation}
E(\bm{\lambda})=\frac{1}{2}\sum\limits_{i=1}^N \lambda_i^2-\frac{1}{2}\sum\limits_{i\neq j}\ln |\lambda_i-\lambda_j|.
\end{equation}
Rescaling the eigenvalues, $\bm{\lambda}=\sqrt{N}\bm{x}$, we obtain
\begin{equation}
E(\bm{x})=\frac{N}{2}\sum\limits_{i=1}^N x_i^2-\frac{1}{2}\sum\limits_{i\neq j}\ln |x_i-x_j|,
\end{equation}
up to a constant shift. We now define the linear statistics in the rescaled variables as
\begin{equation}\label{rescaledA}
\frac{1}{N}\sum_{i=1}^N|x_i|^m=A\,.
\end{equation}
Our goal is to derive the probability density of $A$ which, by definition, can be written as
\begin{equation}
\label{ratio1}
\mathcal{P}(A,m)=\frac{1}{Z_{\beta}}\int\limits_{-\infty}^{\infty}\exp\left[-\beta E(\bm{x})\right]\ \delta\left( \frac{1}{N}\sum_{i=1}^{N} |x_i|^m -  A\right)\prod\limits_{i=1}^N dx_i,
\end{equation}
where
\begin{equation}
Z_{\beta}=\int\limits_{-\infty}^{\infty}\exp\left[-\beta E(\bm{x})\right]\ \prod\limits_{i=1}^N dx_i
\end{equation}
is the partition function of the Coulomb gas, unconstrained by Eq.~(\ref{rescaledA}).  Using the exponential representation of the delta function, we can rewrite Eq.~(\ref{ratio1}), up to an overall constant factor, as
\begin{equation}
\mathcal{P}(A,m)\propto \frac{\int\limits_{-i\infty}^{i\infty}d\Lambda\int\limits_{-\infty}^{\infty}\exp\left[-\beta E(\bm{x},\Lambda)\right]\ \prod\limits_{i=1}^N dx_i}{ Z_{\beta}}\,,
\end{equation}
where
\begin{equation}
E(\bm{x},\Lambda)= \frac{N}{2}\sum\limits_{i=1}^N x_i^2-\frac{1}{2}\sum\limits_{i\neq j}\ln |x_i-x_j| -N^2\Lambda\left( \frac{1}{N}\sum_{i=1}^{N} |x_i|^m - A\right).
\end{equation}
is the constrained energy. Exploiting the large parameter $N\to \infty$, we can reformulate the problem
in a continuum language by introducing the density of rescaled eigenvalues
\begin{equation}
\rho(x)=\frac{1}{N}\sum\limits_{i=1}^N\delta(x-x_i)\,,
\end{equation}
which can be treated as a continuous field. The constrained energy can be expressed as a functional
\begin{multline}
E(\bm{x},\Lambda)=N^2\mathcal{E}[\rho(x),\Lambda] = N^2 \bigg[\frac{1}{2} \int_{-\infty}^{\infty} x^2\,\rho(x)\,dx
-  \\ - \frac{1}{2}\int_{-\infty}^{\infty}\int_{-\infty}^{\infty} \ln|x-y| \rho(x) \rho(y) dx\,dy\,- \Lambda\left(\int_{-\infty}^{\infty} |x|^m \rho(x)\,dx-A\right)\bigg]\,,
\end{multline}
while the multiple integrals over $x_i$ turn into functional integrals over the (normalized to unity) density $\rho(x)$:
\begin{equation}
\mathcal{P}(A,m) \propto \frac{\int \mathcal{D} [\rho] d\Lambda\exp\left[-\beta N^2 \mathcal{E}[\rho(x),\Lambda]-\mathcal{O}(N)\right]\ \delta\left( \int \rho(x)dx -  1\right)}{ \int \mathcal{D} [\rho] \exp\left[-\beta N^2 \mathcal{E}[\rho(x),0]-\mathcal{O}(N)\right]\ \delta\left( \int \rho(x)dx -  1\right)}\,.
\end{equation}
The $\mathcal{O}(N)$ terms in this expression describe contributions from entropy and self-interactions \cite{Dyson1962, DeanMajumdar_2008}. In the large-$N$ limit the energy terms $\mathcal{O}(N^2)$  are dominant, so that we can neglect the $\mathcal{O}(N)$ terms.  Using again the exponential representation of delta functions, we obtain
\begin{equation}
\label{pdf10}
\mathcal{P}(A,m)\propto \frac{\int \mathcal{D} [\rho] d\Lambda d\mu\exp\left[-\beta N^2 \mathcal{E}[\rho(x),\Lambda,\mu]\right]\ }{ \int \mathcal{D} [\rho] d\mu\exp\left[-\beta N^2 \mathcal{E}[\rho(x),0,\mu] \right]\ }\,,
\end{equation}
where the modified energy,
\begin{multline}\label{energy1}
\mathcal{E}[\rho(x),\Lambda,\mu,A] = \frac{1}{2} \int_{-\infty}^{\infty} x^2\,\rho(x)\,dx
- \frac{1}{2}\int_{-\infty}^{\infty}\int_{-\infty}^{\infty} \ln|x-y| \rho(x) \rho(y) dx\,dy\,\\- \Lambda\left(\int_{-\infty}^{\infty} |x|^m \rho(x)\,dx-A\right) - \mu\left(\int_{-\infty}^{\infty}  \rho(x)\,dx-1\right)\,,
\end{multline}
now includes both $\Lambda$ and $\mu$.

At large $N$ the functional integrals in Eq.~(\ref{pdf10}) can be evaluated by the saddle-point method. This brings about a variational problem, where one should minimize the modified energy $\mathcal{E}[\rho(x),\Lambda,\mu]$ over all possible density configurations $\rho(x)$,  and ultimately express parameters $\Lambda$ and $\mu$ (which play the role of Lagrange multipliers) through $A$ by using  the continuum version of Eq.~(\ref{rescaledA}),
\begin{equation}\label{Im}
I_m[\rho(x)]\equiv\int |x|^m \rho(x)\,dx=A\,,
\end{equation}
and the condition that total mass of the gas, $M$, is equal to unity:
\begin{equation}\label{normalization}
M=\int \rho(x)\,dx = 1.
\end{equation}
Once the solution of the minimization problem, $\rho(x)=\rho_A(x)$, $\Lambda=\Lambda_A$ and $\mu=\mu_A$, is found, we can evaluate $\mathcal{P}(A,m)$, up to a pre-exponential factor, as
\begin{equation}
\label{collective}
\mathcal{P}(A,m) \sim  \exp\left[-\beta N^2 \Psi(A,m)\right]
\end{equation}
with a rate function
\begin{equation}\label{R(A,m)}
 \Psi(A,m)=\mathcal{E}[\rho_A(x),A,\mu_A]-\mathcal{E}[\rho_W(x),0,\mu_0]\,.
\end{equation}
Here $\rho_W(x)$ is the density profile unconstrained by $A$: the famous Wigner's semicircular distribution \cite{Wignerpaper}
\begin{equation}
\label{Wigner}
\rho_W(x)=\frac{1}{\pi}\sqrt{2-x^2}\,,
\end{equation}
while $\mu_0$ is the corresponding unconstrained value of $\mu$ \cite{Forrester}. It is evident from Eq.~(\ref{R(A,m)}) that the rate function $\Psi(A,m)$ vanishes when $\rho_A(x)=\rho_W(x)$. In this case $A$ is equal to its (unconstrained) \emph{expected} value
$A=\bar{A}\equiv \bar{A}_m$. We perform the minimization of the modified energy functional (\ref{energy1}) and calculate  the collective rate function $\Psi(A,m)$ in Sec.~\ref{lstat}.

Importantly,  the Wigner's semicircular distribution  (\ref{Wigner})  lives on a compact support, $|x|<\sqrt{2}$.  As we shall see,  the conditioned distributions $\rho_A(x)$ also have compact support. Somewhat surprisingly, however, we find that for $0<m<2$
a single-connected compact support of $\rho_A(x)$ breaks up into two separate compact supports once $A$ exceeds an $m$-dependent critical value.   Such a breakup corresponds to the appearance of a macroscopic gap in the conditioned spectrum of the matrix. In the supercritical region the gap grows with $A$. A qualitatively similar, but quantitatively different, scenario occurs in the Ginibre ensemble (non-Hermitian random matrices with Gaussian entries) and the associated 2D one-component plasma, where the disc-annulus phase transition of the optimal charge density was observed \cite{Cunden_2015,Cunden_2016}\footnote{\label{truncated}The emergence of a gap in the optimal eigenvalue density was also observed in 1D Coulomb gases when studying \emph{truncated} linear statistics, associated with $K<N$ largest eigenvalues of the Gaussian and Laguerre ensembles of random matrices \cite{Grabsch2022, Grabsch2017}.}. We discuss the emergence of the gap and a related phase transition in the limit of $N\to \infty$ in Sections \ref{double} and \ref{order}.
This phase transition  is related to the nature of a singularity of the optimal charge density at $x=0$. The order $p$ of this phase transition is $m$-dependent, and we present analytical and numerical arguments which show  that it is equal to $p=1+1/|m-1|$. In particular, the transition becomes of infinite order for $m=1$.

For $0<m<1$ the critical value of $A$ for the breakup coincides with the expected value of the linear statistics $A=\bar{A}$. In other words, the transition occurs in this case at the ground state of the Coulomb gas, which corresponds to the Wigner's semicircular distribution (\ref{Wigner}).

Crucially,  the collective large-deviation scaling, described by Eq.~(\ref{collective}), does not hold for all values of $A$ and $m$. As we show in Sec. \ref{evaporation}, for $m>2$  the collective scaling can give way to a \emph{single-particle} scaling of the form
\begin{equation}\label{singlescaling2}
\mathcal{P}(A,m) \sim \exp\left[- \frac{\beta}{2} N^{1+\frac{2}{m}}\left(A-\bar{A}\right)^{\frac{2}{m}}\right]\,.
\end{equation}
Here, as well as in Eq.~(\ref{collective}) for the collective rate function, we have neglected
subleading terms -- those with smaller powers of $N$ -- in the exponent.

The single-particle scaling (\ref{singlescaling2}) holds, in the limit of $N\to \infty$, at $A>\bar{A}$ and fixed $A-\bar{A}=\mathcal{O}(1)$. Noticeable in Eq.~(\ref{singlescaling2}) is the factor $N^{1+\frac{2}{m}}$ instead of $N^2$ which appears in Eq.~(\ref{collective}).  At $m>2$ the probability~(\ref{singlescaling2}) is much higher  than the collective one~(\ref{collective}).  Similarly to the breakup transition, this dramatic change of scaling behavior occurs at the ground state of the Coulomb gas. However, in contrast to the breakup, here it has the character of a second-order phase transition for all $m>2$.  The mechanism behind the single-particle scaling (\ref{singlescaling2}) is the ``evaporation" phenomenon, where either the maximum, or the minimum eigenvalue ``escapes" from the rest of the eigenvalues, and it dominates the linear statistics, while the rest of the eigenvalues stay inside the Wigner's semicircle $\rho_W(x)$.
Similar evaporation phenomena have been encountered in problems of statistics of the maximum or minimum eigenvalue of random matrices \cite{Majumdar_2014,MajumdarVergassola,Cunden_2016}.  To our knowledge, the present work is the first one to have observed the evaporation scenario for a linear statistics of Gaussian random matrices\footnote{\label{Texier}A  change of scaling from collective to single-particle was previously found for the particular linear statistics $\sum_i^N x_i^{-1}$  of a Coulomb gas confined by the potential $V(x)=x-\ln x$ at $x>0$ \cite{Texier_2013}. However, the nature of the phase transition in these two systems is different.}.

By its very nature, the present work deals with rare events that are virtually impossible to sample by means of conventional Monte-Carlo simulations. Therefore, we verified our main theoretical predictions by using the multicanonical sampling of GOE matrices ($\beta=1$) \cite{Saito2010}, based on the Wang-Landau algorithm \cite{Wang-Landau}.

\begin{figure} [ht]
\includegraphics[width=0.3\textwidth,clip=]{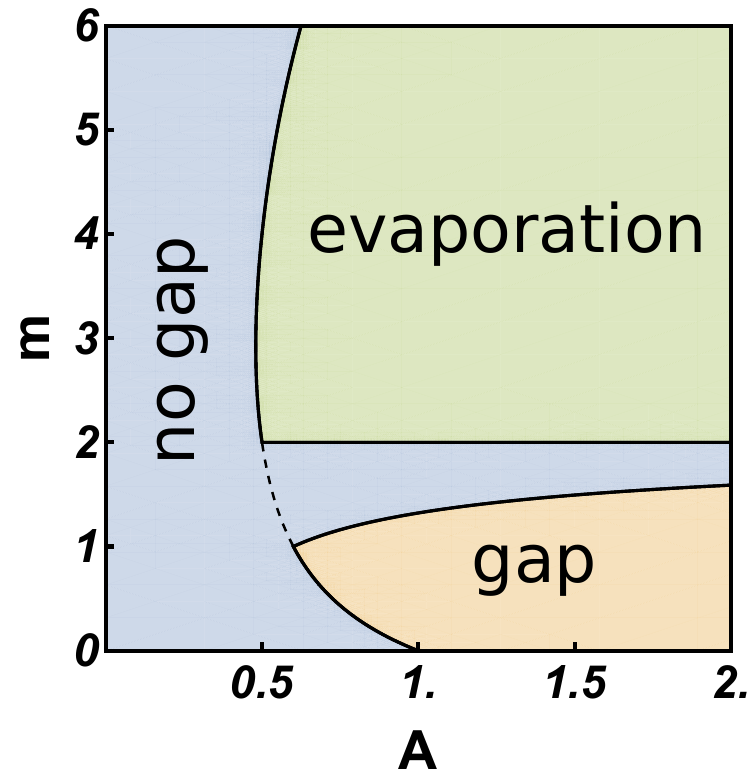}
\caption{Phase diagram of the large-deviation statistics (\ref{exactsum}) on the $(A,m)$ plane. The curved line corresponds to the ground state of the system $A=\bar{A}$, when the Coulomb gas density is described by the Wigner's semicircular distribution~(\ref{Wigner}).}
\label{Fig:phasediagram}
\end{figure}

We shall present the calculation of the single-particle rate function $\Psi_s(A,m)$ in Sec. \ref{evaporation}.  Our main results are briefly summarized and discussed  in Sec. \ref{discussion}. Figure \ref{Fig:phasediagram} shows a phase diagram of this system on the $(A,m)$ plane.

\section{Collective scaling of rate function}
\label{lstat}

The minimization of the modified energy (\ref{energy1}) brings about three conditions \cite{GrabschTexier2016}. One of them,
\begin{equation} \label{Variation}
 \frac{\delta \mathcal{E}[\rho(x),\Lambda, \mu,A]}{\delta \rho(x)}=0\,,
\end{equation}
yields a linear integral equation for $\rho(x)$ which describes the mechanical equilibrium of charges in an effective potential $V_{\text{eff}}(x,\Lambda)$:
\begin{equation}\label{inteq2}
\int_{-\infty}^{\infty} \ln|x-y| \,\rho(y)\,dy =V_{\text{eff}}(x,\Lambda) -\mu \equiv \frac{1}{2} x^2 -\Lambda |x|^m -\mu\,.
\end{equation}
This equation holds in the region(s) of space where $\rho(x)>0$. The other two conditions are
\begin{equation} \label{Variation2}
\frac{\partial \mathcal{E}[\rho(x),\Lambda, \mu,A]}{\partial \Lambda}=0\qquad\text{and}\qquad \frac{\partial \mathcal{E}[\rho(x),\Lambda, \mu,A]}{\partial \mu}=0\,.
\end{equation}
The first of them imposes the condition (\ref{Im}) and yields a relationship between $\Lambda$ and  $A$, whereas the second one imposes the normalization condition~(\ref{normalization}).

It is impossible to satisfy Eq.~(\ref{inteq2}) if $\rho(x)>0$ on whole line $|x|<\infty$, because otherwise the l.h.s. and r.h.s. would have different asymptotics at $|x|\rightarrow\infty$. (In the particular case of $\Lambda=0$, which corresponds to the Wigner's semicircle, this fact is well known, see \textit{e.g.} Ref. \cite{Forrester}.) Therefore,  the solution of Eq.~(\ref{inteq2}) must have finite support. Once we have realized it, we should add one more condition to Eq. (\ref{Variation}), which follows from variation of the \textit{a priori} unknown  positions of the boundaries of the compact support:
\begin{equation}\label{BC}
\rho = 0\quad \text{at support boundaries\,.}
\end{equation}
Equation~(\ref{inteq2}), subject to the normalization condition~(\ref{normalization}), is equivalent to the equation
\begin{equation}\label{force}
\text{p.v.}\!\int \frac{\rho(y)}{x-y}\,dy =  x -\Lambda \, m\,|x|^{m-1} \text{sgn} x\,,
\end{equation}
subject to the same normalization condition.  Equation~(\ref{force}) can be obtained by differentiating Eq.~(\ref{inteq2}) with respect to $x$. It describes the force balance of the charge at point $x$: the l.h.s is the net force from the other charges, while the r.h.s. is the force exerted by the effective potential $V_{\text{eff}}(x,\Lambda)$.

It will be useful for the following to consider a more general family of solutions $\rho(x)$ of Eq.~(\ref{inteq2}) [or, equivalently, of Eq.~(\ref{force})] which are not subject to the normalization condition (\ref{normalization}) and can have an arbitrary mass $M$. These solutions can be parameterized by $\Lambda$ and $\mu$, and they have an important scale-invariance property. Suppose that $\rho(x)$ is such a solution. Then the corresponding mass, moment, and energy of the Coulomb gas are
\begin{eqnarray}
  M(\Lambda,\mu) &=& \int \rho(x) dx\,,\label{Mgen} \\
   A(\Lambda,\mu) &=& \int |x|^m \rho(x) dx \,,\label{Agen} \\
  \mathcal{E}(\Lambda,\mu) &=& \frac{1}{2} \int x^2\,\rho(x)\,dx\,
- \frac{1}{2}\iint \ln|x-y| \rho(x) \rho(y) dx\,dy\, =\frac{I_2+2\Lambda A +2\mu}{4}\,, \label{Egen}
\end{eqnarray}
where the last equality is a consequence of Eq.~(\ref{inteq2}), and $I_m$ is defined in Eq.~(\ref{Im}).
Introducing the rescaling
\begin{equation}
    \tilde{x}=\alpha x\qquad \text{and}\qquad \tilde{\rho}(\tilde{x})=\alpha^{-1}\rho(\alpha x),
    \label{scaleinvariance}
\end{equation}
we can see that $\tilde{\rho}$ solves Eq.~(\ref{inteq2}) with rescaled values of $\Lambda$ and $\mu$:
\begin{equation}
    \Lambda\rightarrow\tilde{\Lambda}=\alpha^{m-3}\Lambda \quad \text{and}\quad \mu\rightarrow\tilde{\mu}=\alpha^{-2}\mu+M(\Lambda,\mu)\ln\alpha
\end{equation}
The rescaled mass, moment and energy are
\begin{equation} \label{P010}
    \tilde{M}(\tilde{\Lambda},\tilde{\mu})=\alpha^{-2}M(\Lambda,\mu)\,,\qquad \tilde{A}(\tilde{\Lambda},\tilde{\mu})=\alpha^{-m-2}A(\Lambda,\mu)\,,\qquad \tilde{\mathcal{E}}(\tilde{\Lambda},\tilde{\mu})=\alpha^{-4}\mathcal{E}(\Lambda,\mu)\,.
\end{equation}

\begin{figure} [ht]
\includegraphics[width=0.4\textwidth,clip=]{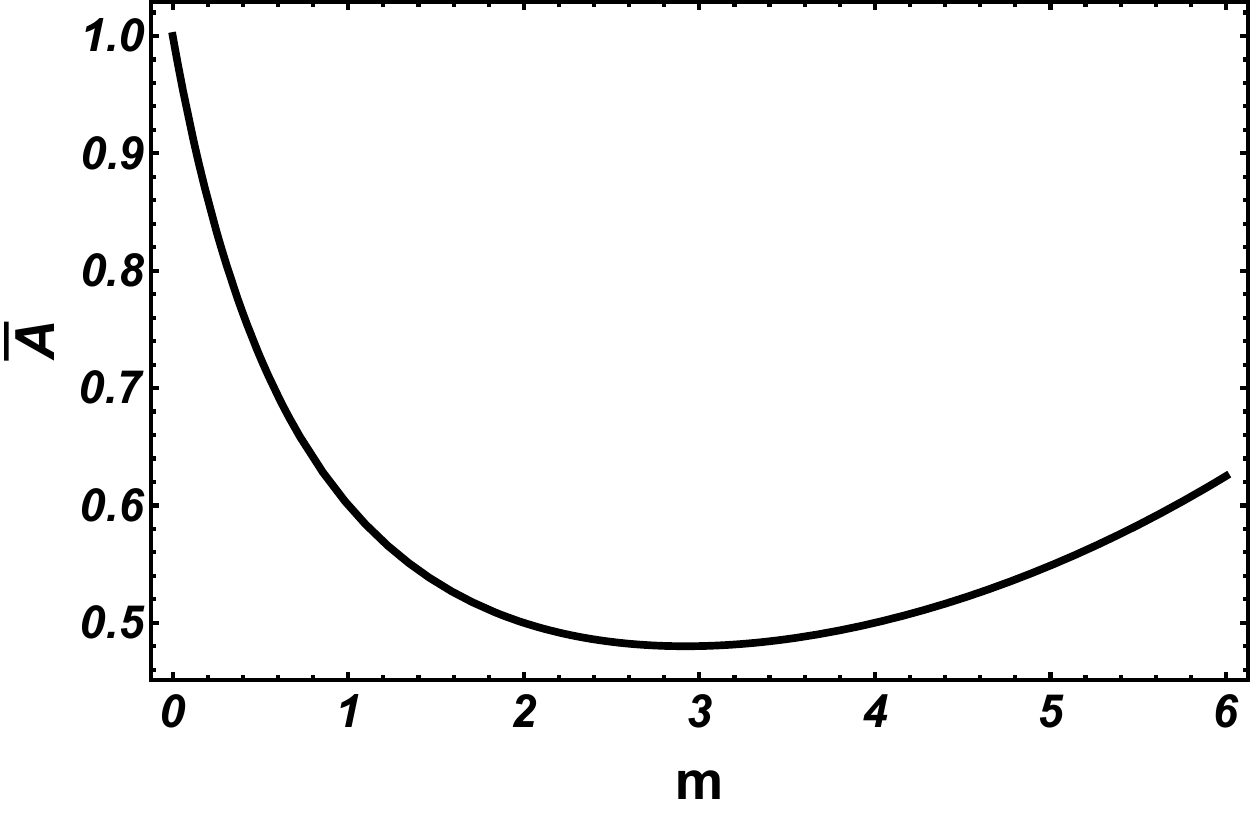}
\caption{Expected value $\bar{A}$ of the linear statistics as a function of $m$.}
\label{Fig:barA}
\end{figure}

Now let us return to Eq.~(\ref{inteq2}) subject to the normalization condition (\ref{normalization}). Before dealing with arbitrary $A$, we note that the Wigner's distribution  (\ref{Wigner})  follows from these equations in the absence of any constraint on $A$, that is for $\Lambda=0$.  In order to see it, one should assume that $\rho(x)$  has a single-connected support $x \in (-\sqrt{B},\sqrt{B})$ [where $B$ is to be found alongside with $\rho(x)$] and solve the Carleman's equation~(\ref{inteq2}).  For $\sqrt{B}\neq 2$ the solution is  \cite{Polyanin}
\begin{equation}\label{Carleman}
    \rho(x)=\frac{1}{\pi^2\sqrt{B-x^2}}
    \left[\int\limits_{-\sqrt{B}}^{\sqrt{B}}\frac{\sqrt{B-t^2}V_{\text{eff}}'(t,\Lambda)}{t-x}dt
    +\frac{1}{\ln\frac{\sqrt{B}}{2}}
    \int\limits_{-\sqrt{B}}^{\sqrt{B}}\frac{V_{\text{eff}}(t,\Lambda)-\mu}{\sqrt{B-t^2}}dt\right]\,,
\end{equation}
where the effective potential is $V_{\text{eff}}(x,\Lambda=0)= x^2/2$. The first integral  in Eq.~(\ref{Carleman}), and similar integrals in the following, are understood as Cauchy principal value integrals.  The integrations can be performed explicitly, and we obtain a two-parameter solution:
\begin{equation}
\rho(x; B, \mu) = \frac{\frac{B-4 \mu }{\ln
   \frac{\sqrt{B}}{2}}+2B-4
   x^2}{4 \pi  \sqrt{B-x^2}}\,.
   \label{100}
\end{equation}
By virtue of the boundary condition (\ref{BC}), the numerator in Eq.~(\ref{100}) must vanish at $x=\pm \sqrt{B}$. This condition eliminates one of the parameters, say $\mu$ in favor of $B$, and we arrive at a one-parameter solution:
\begin{equation}
   \mu= \frac{B}{4} \left(1-  \ln \frac{B}{4}\right) \qquad \text{and}\qquad  \rho(x;B)=
   \frac{1}{\pi} \sqrt{B-x^2}.
\end{equation}
The remaining parameter $B$ is determined from the normalization condition (\ref{normalization}).
This gives $B=2$ leading to the Wigner's semicircular distribution (\ref{Wigner}).
The  expected value $\bar{A}=\bar{A}(m)$ of the linear statistics, corresponding to the unconstrained density $\rho_{0}(x)$, is
\begin{equation}\label{barA}
    \bar{A}=\int\limits_{-\sqrt{2}}^{\sqrt{2}}|x|^m \rho_{0}(x)dx=\frac{2^{m/2} \Gamma \left(\frac{m+1}{2}\right)}{\sqrt{\pi } \Gamma \left(\frac{m}{2}+2\right)}\,,
\end{equation}
where $\Gamma(\dots)$ is the gamma function. The dependence of $\bar{A}$ on $m$ is shown in Fig. \ref{Fig:barA}.

\subsection{No spectral gap}
\label{single}

Let us consider the full Eq.~(\ref{inteq2}), which is conditioned on $A$ via the Lagrange multiplier $\Lambda$, and again assume that the solution $\rho_A(x)$ of Eq.~(\ref{inteq2}) is defined on a single-connected  compact support $|x|<\sqrt{B}$.  For $\sqrt{B}\neq 2$  the solution \cite{Polyanin} can be written as
\begin{equation}
    \rho_A(x;B,\mu,\Lambda)=\frac{\Lambda  B^{m/2} \Gamma \left(\frac{m+1}{2}\right) \left\{\frac{B m \, \Re[_2\tilde{F}_1\left(1,\frac{m+1}{2};\frac{m+4}{2};\frac{B}{x^2}\right)]}{x^2}+\frac{4}{\ln \frac{4}{B} \Gamma \left(\frac{m}{2}+1\right)}\right\}+\frac{\sqrt{\pi } \left[\left(B-2 x^2\right) \ln \frac{B}{4}+B-4 \mu \right]}{\ln \frac{B}{4}}}{2 \pi ^{3/2} \sqrt{B-x^2}}
\end{equation}
where $\Re[_2\tilde{F}_1\left(1,\frac{m+1}{2};\frac{m+4}{2};\frac{B}{x^2}\right)]$ is the real part of the regularized hypergeometric function $_2\tilde{F}_1(a,b,c;z)$. As in the case of unconstrained density, we can get rid of the extra parameters by using the constraints (\ref{Im}), (\ref{normalization}) and (\ref{BC}). Equation~(\ref{normalization}) yields
\begin{equation}\label{mu}
\int\limits_{-\sqrt{B}}^{\sqrt{B}}\rho_A(x;B,\mu,\Lambda)dx=-\frac{2 \Lambda  B^{m/2} \Gamma \left(\frac{m+1}{2}\right)}{\sqrt{\pi } \ln \frac{B}{4} \Gamma \left(\frac{m}{2}+1\right)}+\frac{B-4\mu}{2\ln \frac{B}{4}} =1.
\end{equation}
After some algebra, we can eliminate $\mu$. Then Eq.~(\ref{Im}) yields
\begin{equation}\label{Lambda}
   \int\limits_{-\sqrt{B}}^{\sqrt{B}}|x|^m\rho_A(x;B,\Lambda)dx= \frac{2 \Lambda  B^m \Gamma \left(\frac{m+1}{2}\right)^2}{\pi  m \Gamma \left(\frac{m}{2}\right)^2}-\frac{[(B-2) m-4] B^{m/2} \Gamma \left(\frac{m+1}{2}\right)}{4 \sqrt{\pi } \Gamma \left(\frac{m}{2}+2\right)}=A.
\end{equation}
Finally, the condition $\rho(|x| = \sqrt{B})=0$ gives
\begin{equation}\label{1dom}
    \left.\rho_A(x;B)\right|_{x=\pm\sqrt{B}}=\frac{4 \sqrt{\pi } A B^{-\frac{m}{2}} \Gamma \left(\frac{m}{2}+1\right)}{\Gamma \left(\frac{m+1}{2}\right)}-\frac{4 B}{m+2}+B-2=0.
\end{equation}
As this equation for $B$ is implicit, it is convenient to represent the solution in the parametric form with $B$ as the single parameter:
\begin{equation}
\rho_A(x;B)=\frac{(m+2) x^2 \left(B-2 x^2+2\right)+(B-2) B \, \Re[_2F_1\left(1,\frac{m+1}{2};\frac{m+4}{2};\frac{B}{x^2}\right)]}{2 \pi  (m+2) x^2 \sqrt{B-x^2}}\,.
\label{rho_s}
\end{equation}
The rest of parameters $A$, $\Lambda$, and $\mu$ can  be expressed in terms of $B$ with the help of Eqs. (\ref{mu}) and (\ref{Lambda}):
\begin{eqnarray}
  A(B)&=& \frac{B^{m/2} \left[2(m+2)-B (m-2)\right] \Gamma \left(\frac{m+1}{2}\right)}{4 \sqrt{\pi } (m+2) \Gamma \left(\frac{m}{2}+1\right)}\,, \label{AA(B)} \\
  \Lambda(B) &=& \frac{\sqrt{\pi } (B-2) B^{-\frac{m}{2}} \Gamma \left(\frac{m}{2}\right)}{4 \Gamma \left(\frac{m+1}{2}\right)}\,, \label{Lambda(B)} \\
  \mu(B)&=& \frac{1}{2} \left(\frac{2-B}{m}+\frac{B}{2}+\ln \frac{4}{B}\right)\,. \label{mu(B)}
\end{eqnarray}

\begin{figure}[H]
\centering
\includegraphics[scale=0.5]{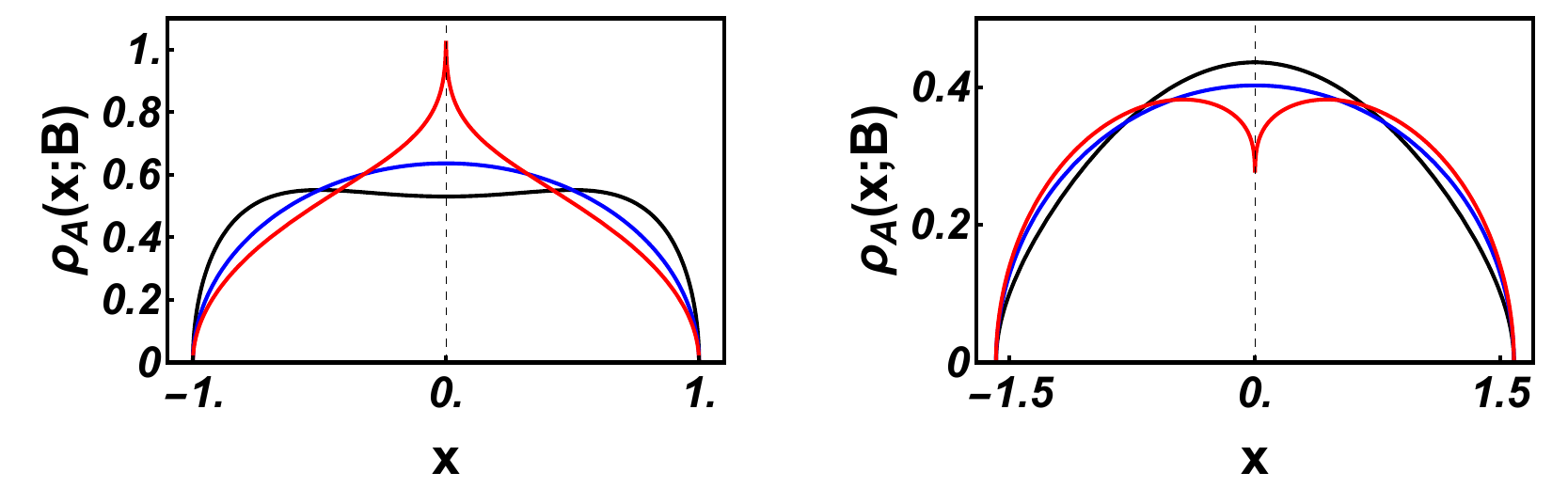}
\caption{Optimal density profiles $\rho_A(x;B)$ as described by Eq.~(\ref{rho_s}). The left panel corresponds to $B=1$, for which $A<\bar{A}$, where $\bar{A}$ is given by Eq.~(\ref{barA}). The shown examples are: $m=4$ so that $A\simeq 0.16$ and  $\bar{A}=0.5$ (black); $m=2$ so that $A\simeq 0.25$ and $\bar{A}= 0.5$ (blue), and $m=5/4$ so that $A\simeq 0.33$ and  $\bar{A}\simeq 0.56$ (red). The right panel corresponds to $B=2.5$, for which $A>\bar{A}$. The shown examples are $m=6$ so that $A\simeq 0.68$ and $\bar{A}=0.5$ (black),
$m=2$ so that $A\simeq 0.63$ and $\bar{A}= 0.5$ (blue), and $m=5/4$ so that $A\simeq 0.68$ and $\bar{A}\simeq 0.56$ (red).}
\label{Fig:fig01}
\end{figure}

Several examples of the optimal density for different values of $B$ and $m$ are plotted in Fig. \ref{Fig:fig01}. To remind the reader, (i) $B=2$ corresponds to the semicircular Wigner distribution (\ref{Wigner}), (ii) for any $m$ we have $A(B=2)=\bar{A}$ as described by Eq.~(\ref{barA}), and (iii) $\Lambda(B= 2)=0$.  For $m=2$ one obtains a rescaled Wigner distribution $\rho_A(x;B)=\frac{2 \sqrt{B-x^2}}{\pi B}$. This is an example of the scale-invariance property (\ref{scaleinvariance}) with $\alpha=\sqrt{1-2\Lambda}=\sqrt{\frac{2}{B}}$. The rescaled effective potential Eq. (\ref{inteq2}) is $V_{\text{eff}}=(1/2)\tilde{x}^2$ leading to the Wigner semicircle. $A<\bar{A}$ corresponds to $\Lambda<0$ and to ``squeezed" distributions. Indeed, the effective potential $V_{\text{eff}}(x,\Lambda)$  confines the gas stronger as $\Lambda$ decreases, and vice versa.

\begin{figure}[H]
\centering
\includegraphics[scale=0.4]{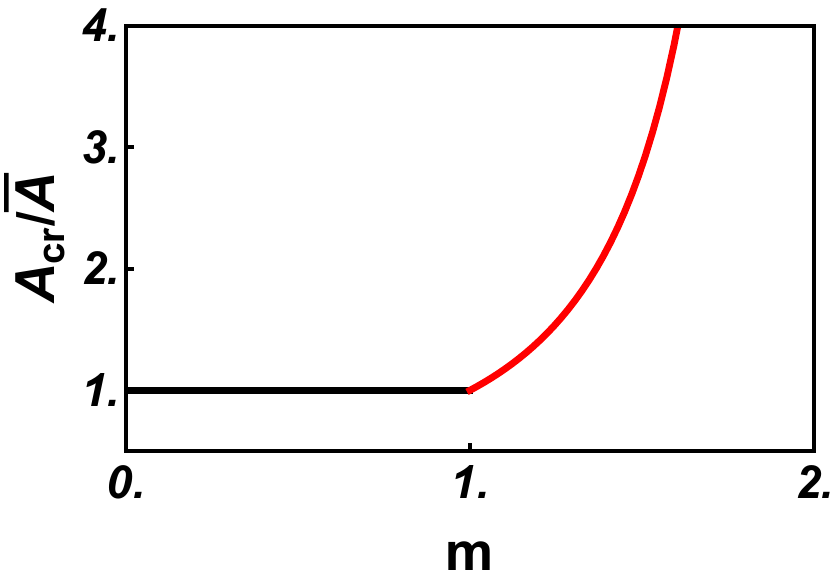}
\caption{Rescaled critical value of the linear statistics $A_{\text{cr}}/\bar{A}$, see Eq.~(\ref{Acr}), for the transition from a single- to a double-interval optimal density, as a function of $m$.}
\label{Fig:A_mu}
\end{figure}

Noticeable in Fig.~\ref{Fig:fig01} are a positive (at $A<\bar{A}$) or negative (at $A>\bar{A}$) peak of the optimal density~(\ref{rho_s}) at $x=0$. The asymptotic behavior of $\rho_A(x; B)$ in the vicinity of $x=0$ is the following:
\begin{equation}\label{rho_exp}
\rho(x\to 0;B)\,{=}\, \begin{cases}
    \frac{B (m-2)+2 m}{2 \pi  \sqrt{B} (m-1)}-\frac{ m  }{\pi } \tan \left(\frac{\pi  m}{2}\right) \Lambda |x|^{m-1}+\ldots \qquad \text{for $m\neq 1$}, \\
    \frac{\sqrt{B}}{\pi }-\frac{\Lambda  }{\pi ^2}\ln \frac{4 B}{x^2}+\ldots \qquad \text{for $m=1$}\,.
\end{cases}
\end{equation}
That is, the  single-interval solution~(\ref{rho_s}) develops a  power-law singularity (for $0<m<1$) or a logarithmic singularity (for $m=1$)  at the origin. The character of the singularities (\ref{rho_exp}) is such that the solution for $0<m\leq 1$ becomes negative -- and therefore inadmissible -- for any $\Lambda>0$. For $1<m<2$ there is no singularity at $x=0$, but the leading term in the expansion (\ref{rho_exp}) becomes negative when $B$ exceeds a critical value $B_{\text{cr}}>0$, so the single-interval solution breaks down here as well. For all $m$ the critical value $B_{\text{cr}}>0$ is the minimum value of $B$ at which the solution vanishes at $x=0$. Using Eq.~(\ref{AA(B)}), we can also find the corresponding critical value of $A$, so that the single-interval solution breaks down at $A>A_{\text{cr}}$. The critical values of $B$ and $A$ are
\begin{equation}\label{Acr}
  B_{\text{cr}}=    \begin{cases}
       \quad 2\qquad \text{for $m\leq 1$}, \\
        \frac{2m}{2-m}\qquad \text{for $1<m<2$}\,.
    \end{cases}
 A_{\text{cr}}=    \begin{cases}
        \bar{A}=\frac{2^{m/2} \Gamma \left(\frac{m+1}{2}\right)}{\sqrt{\pi } \Gamma \left(\frac{m}{2}+2\right)}\qquad \text{for $m\leq 1$}, \\
        \frac{2^{m/2} \left(\frac{m}{2-m}\right)^{m/2} \Gamma \left(\frac{m+3}{2}\right)}{\sqrt{\pi } \Gamma \left(\frac{m}{2}+2\right)}\qquad \text{for $1<m<2$};
    \end{cases}\qquad
\end{equation}
As we found, at $B>B_{\text{cr}}$ and $0<m<2$ the single-interval solution (\ref{rho_s}) breaks down, and the true optimal density of the Coulomb gas
splits into two domains, separated by a gap around $x=0$. Remarkably, for $0<m\leq 1$,
this transition occurs  already in the ground state of the system which corresponds to the Wigner semicircle and to $A=\bar{A}$, see Fig. \ref{Fig:A_mu}.

\begin{figure}[H]
\centering
\includegraphics[scale=0.4]{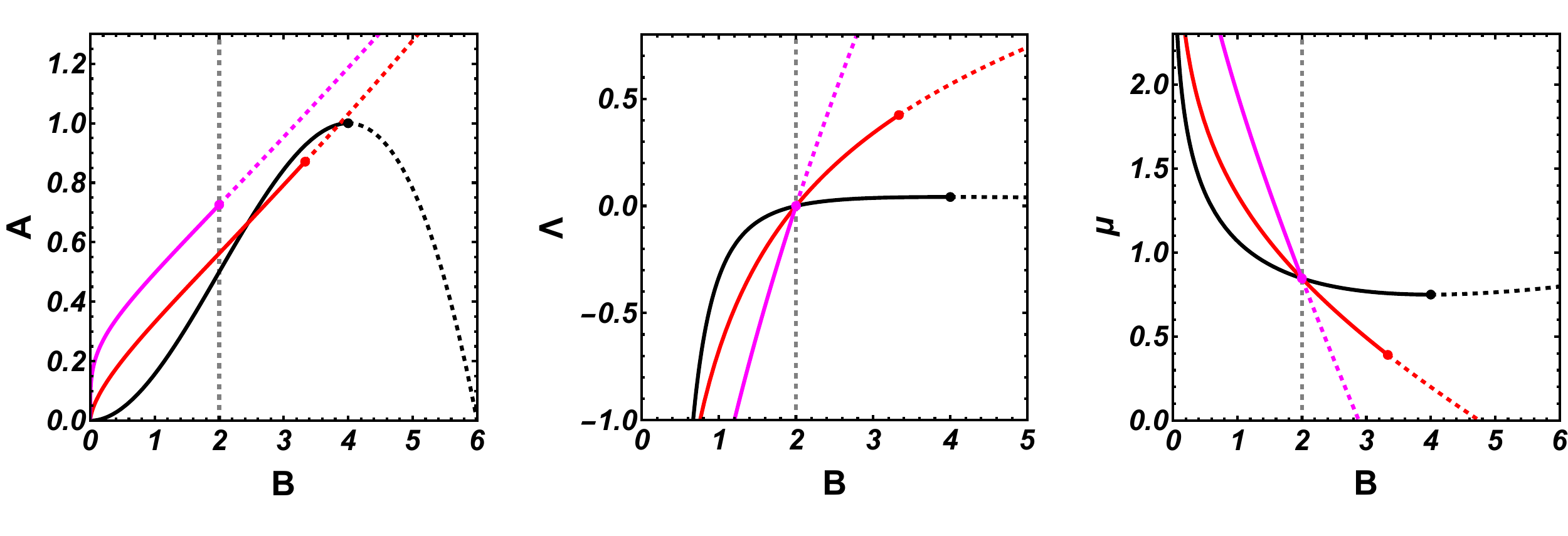}
\caption{The linear statistics $A=A(B)$ and the Lagrange multipliers $\Lambda=\Lambda(B)$ and $\mu=\mu(B)$, see Eqs.~(\ref{AA(B)})-(\ref{mu(B)}), for different values of $m$:  $m=4$ (black), $m=5/4$ (red), and $m=1/2$ (magenta). The solid curves correspond to the physically relevant regions $B\leq B_{*}$ (for $m=4$) and $B\leq B_{\text{cr}}$ (for $m=5/4$ and $1/2$).}
\label{Fig:ALM}
\end{figure}

For $m>2$  the single-interval solution (\ref{rho_s}) continues to be formally valid for all $B<B_*=\frac{2m}{m-2}$\footnote{\label{threedomain}At $B>B_*$ the gas density (\ref{rho_s}) becomes negative near  the edges of support of the solution. This hints at a topological transition from a single-interval optimal configuration to a three-interval one.}.
According to Eq.~(\ref{AA(B)}), the corresponding critical value of $A$ is
\begin{equation}\label{Astar}
A_*=\frac{\left(\frac{2 m}{m-2}\right)^{m/2} \Gamma \left(\frac{m+1}{2}\right)}{2 \sqrt{\pi }\, \Gamma \left(\frac{m}{2}+2\right)}\,.
\end{equation}
However, at $N\to \infty$, the true minimum of the action here, already at $\Lambda>0$, that is $A>\bar{A}$, is reached via  one-particle evaporation. We will analyze these regimes, and the corresponding phase transitions, in Sections \ref{double} and \ref{evaporation}, respectively. In the remainder of this Section we will complete the solution of the problem for $B \leq B_*$ by calculating the rate function $\Psi(A)$ which determines the probability distribution~(\ref{collective}). Before doing that, we plot in Fig. \ref{Fig:ALM} the functions $A(B)$, $\Lambda(B)$ and $\mu(B)$, as described by Eqs.~(\ref{AA(B)})-(\ref{mu(B)}),  for several values of $m$. The solid curves correspond to the physically relevant regions $B \leq B_{\text{cr}}$ or $B \leq B_{*}$ .

Our results for $A=A(B)$ and $\Lambda=\Lambda(B)$ in Eqs.~(\ref{AA(B)}) and (\ref{Lambda(B)}), respectively, make it possible to calculate the rate function $\Psi_1(B)$ in a parametric form (we use the subscript $1$ to indicate single-interval support) by employing the ``shortcut relation" $d\Psi_1/dA = \Lambda$, see \textit{e.g.} Ref. \cite{Cunden2016}. We can write $d\Psi_1/dA=(d\Psi_1/dB)(dB/dA)$ and obtain the equation
\begin{equation}
\frac{d\Psi_1}{dB} = \Lambda(B) \frac{dA(B)}{dB}\,.
\label{shortcut1}
\end{equation}
Using Eq. (\ref{AA(B)}) for $A(B)$ and integrating Eq.~(\ref{shortcut1}) from $B=2$ to $B$ , we obtain
\begin{equation}
\Psi_1(B)=\frac{1}{4}\ln \frac{2}{B}+\frac{(2-B) [B (m-2)-6 m+4]}{32 m}.
\label{k_RateB}
\end{equation}
The same result  (\ref{k_RateB})  can be also obtained from Eq.~(\ref{R(A,m)}) by evaluating the total energy of the Coulomb gas in Eq.~\eqref{energy1}, or by substituting the second moment,
\begin{equation}\label{I2m}
I_2 =  \int\limits_{-\sqrt{B}}^{\sqrt{B}}|x|^2\rho_A(x;B)dx=\frac{B \left[4 m-B (m-2)\right]}{8 (m+2)},
\end{equation}
into the r.h.s. of Eq.~(\ref{Egen}). In both cases one should subtract the energy, corresponding to the unconstrained Wigner distribution,  $\mathcal{E}[\rho_W(x)] =\frac{1}{8}\left(3+\ln 4\right) \simeq 0.54828\dots$.

\begin{figure}[H]
\centering
\includegraphics[width=.35\linewidth]{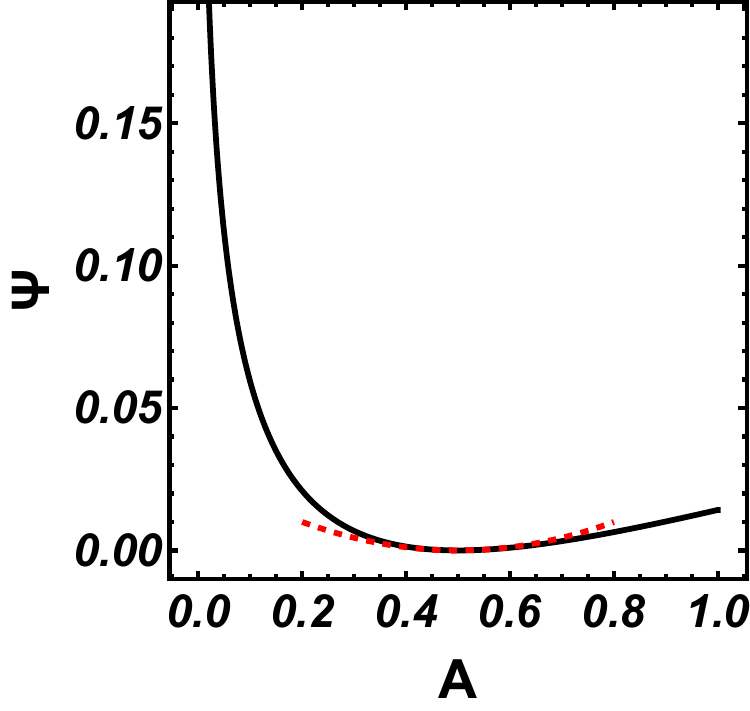}
     \caption{The collective rate function $\Psi_1(A)$ for $m=4$ (black). The red dashed line shows the Gaussian asymptotic (\ref{R(A)_gauss}).}\label{Fig:Data1}
\end{figure}

Equation (\ref{k_RateB}) alongside with $A=A(B)$ from Eq.~(\ref{AA(B)}) give $\Psi_1=\Psi_1(A)$ in the parametric form. The plot of $\Psi_1(A)$ for a particular value of $m=4$ is shown in Fig. \ref{Fig:Data1}. As to be expected on the physical grounds, the minimum of the rate function is at $A=\bar{A}$.  A Taylor expansion of $\Psi_1(A)$  in the vicinity of $A=\bar{A}$ yields a quadratic asymptotic
\begin{equation}
\Psi_G(A)\simeq\frac{ 2^{2-m} \pi  \Gamma^2 \left(\frac{m}{2}+2\right) \left[A-\frac{2^{m/2} \Gamma \left(\frac{m+1}{2}\right)}{\sqrt{\pi } \Gamma \left(\frac{m}{2}+2\right)}\right]^2}{m (m+2)^2 \Gamma^2 \left(\frac{m+1}{2}\right)}\,.
\label{R(A)_gauss}
\end{equation}
Alongside with Eq.~(\ref{collective}), this asymptotic describes typical, Gaussian fluctuations of $A$ around the mean with the variance $N^{-2}\sigma^2$, where
\begin{equation}\label{variance}
\sigma^2=\frac{m (m+2)^2 \Gamma^2 \left(\frac{m+1}{2}\right)}{ 2^{3-m} \pi \Gamma^2 \left(\frac{m}{2}+2\right)}.
\end{equation}
For even integer values of $m$ this variance coincides with the one obtained in Ref. \cite{Cunden}, where typical fluctuations of the quantity $\sum_{i=1}^N \lambda_i^m$ (without the absolute value) at large $N$ were studied.

For $m=2$, the collective rate function, determined by Eqs. (\ref{AA(B)}) and (\ref{k_RateB}), can be easily expressed in terms of $A$:
\begin{equation}
    \Psi_1(A)=\frac{2A-1-\ln (2A)}{4}\,.
\end{equation}
When multiplied by $\beta=2$, this rate function coincides with the one describing large-deviation statistics of the  potential energy,  $(1/2)\sum_{i=1}^N x_i^2$, of $N\gg 1$ non-interacting fermions in a harmonic trap \cite{Grela}. The latter rate function arises from the large-$N$ asymptotic of a gamma distribution \cite{Grela}.

\subsection{Spectral gap}
\label{double}

To determine the optimal density and the rate function in the regime with a gap, that is at $A>A_{\text{cr}}$, we have to solve the double-interval integral equation
\begin{equation}\label{integn}
\int\limits_{-\sqrt{B}}^{-\sqrt{b}}\rho(y)\ln|x-y|dy+\int\limits_{\sqrt{b}}^{\sqrt{B}}\rho(y)\ln|x-y|dy= \frac{x^2}{2}-\Lambda |x|^{m} - \mu \qquad \mbox{at~~} b<x^2<B\,,
\end{equation}
where $B>b>0$, $\rho(x)$ is normalized to unity, and the parameters $B$ and $b$ are \textit{a priori} unknown. Let us rescale $x/\sqrt{B} = \tilde{x}$, denote $\sqrt{b/B}=k>0$, and introduce a rescaled density $\tilde{\rho}(\tilde{x}) = \rho(\sqrt{B} \tilde{x})/\sqrt{B}$, so that
\begin{equation}\label{normnew}
\int_{-1}^{-k} \tilde{\rho}(\tilde{x})\,d\tilde{x} + \int_k^1 \tilde{\rho}(\tilde{x})\,d\tilde{x} =1\,.
\end{equation}
In the rescaled variables Eq.~(\ref{integn}) becomes
\begin{equation}\label{DI solution}
\int\limits_{-1}^{-k}\rho(y)\ln|x-y|dy+\int\limits_{k}^{1}\rho(y)\ln|x-y|dy=f(x)\equiv B\left(\frac{x^2}{2}-\lambda |x|^{m} - \mu\right),
\end{equation}
where $k<|x|<1$ and $0<k<1$, $\lambda = \Lambda \,B^{\frac{m}{2}-1}$, $\tilde{\mu} = (\mu+\ln \sqrt{B})/B$, and we dropped the tildes. The solution, according to Ref. \cite{Gautesen}, can be written in the following form:
\begin{multline}
\pi^2 \sqrt{R(x)}  \text{sgn}(x)\rho(x)=\int\limits_{-1}^{-k}+\int\limits_{k}^{1}\frac{\sqrt{R(t)} \text{sgn}(t) f'(t) dt}{t-x} +
\int\limits_{-1}^{-k}+\int\limits_{k}^{1}\bigg(1-t^2-\frac{E(k)}{K(k)}-\frac{x t}{\ln(2/k')}\bigg)\frac{ \text{sgn}(t) f(t) dt}{\sqrt{R(t)}}\,.
\label{sol1}
\end{multline}
Here the sum of integrals over the two symmetric intervals is denoted for brevity by $\int_{-1}^{-k}+\int_{k}^{1}$. Also, $R(x)=(1-x^2)(x^2-k^2)$,  $E(k)$, $K(k)$ are the complete elliptic integrals,  and $k'=\sqrt{1-k^2}$. A simplification comes from the fact that $f(x)$, and  as a result $\rho(x)$, are even functions.  Using this property, we can rewrite Eq.~(\ref{sol1}) as
\begin{equation}
\pi^2 \sqrt{R(x)}  \text{sgn}(x)\rho(x)=\int\limits_{-1}^{-k}+\int\limits_{k}^{1}\frac{\sqrt{R(t)} \text{sgn}(t) f'(t) dt}{t-x} + c B x\,,
\label{sol3}
\end{equation}
where
\begin{equation}\label{defc}
c= -\frac{1}{B\ln\frac{2}{k'}} \int\limits_{-1}^{-k}+\int\limits_{k}^{1} \frac{t \text{sgn}(t) f(t) dt}{\sqrt{R(t)}}\,.
\end{equation}
We were able to perform  the integration in the r.h.s. of Eq.~(\ref{sol3}) analytically (with the help of ``Mathematica") only for $m=1$. This calculation, presented in the next subsection, leads to the prediction of infinite-order phase transition  at $A=A_{\text{cr}}$ in this special case. For other $0<m<2$ we present a non-rigorous perturbative argument, which is based on the asymptotic behavior (\ref{rho_exp}) of the single-interval solution $\rho_A(x;B)$ in the vicinity of $x=0$. It predicts a phase transition at $A=A_{\text{cr}}$ with an $m$-dependent and, in general, fractional order. Where possible, we will support these predictions by numerical simulations.

\subsubsection{$m=1$: general}
\label{m1general}

\begin{figure}[H]
   \begin{minipage}{0.48\textwidth}
     \centering
      \includegraphics[width=.9\linewidth]{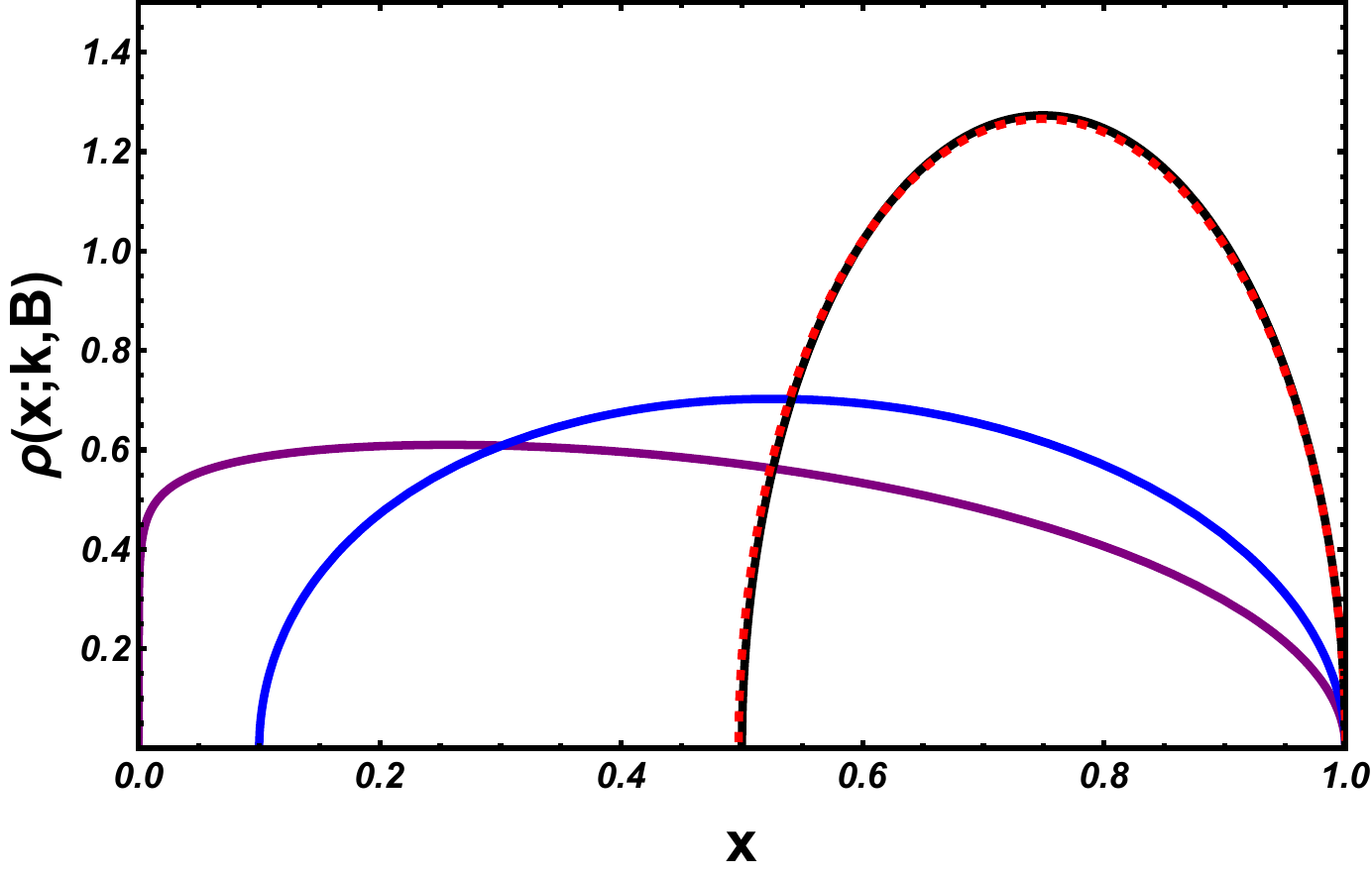}
        \caption{The optimal densities $\rho(x;k,B)$ for the two-interval solution with $m=1$ for the following parameters: $k=10^{-6}$ so that $B\simeq2.30$ and $A\simeq0.67$ (purple);  $k=10^{-1}$ so that $B\simeq4.35$ and $A\simeq1.13$ (blue), and $k=5\cdot 10^{-1}$ so that $B\simeq15.77$ and  $A\simeq2.98$ (black). The red dashed curve corresponds to the rescaled Wigner's semicircular distribution $(A+1)\rho[x(A+1);A]$ (\ref{DI_sc}) for $A\simeq 2.98$. Only the positive half-line is shown.}
\label{Fig:Dens}
   \end{minipage}\hfill
   \begin{minipage}{0.48\textwidth}
     \centering
     \includegraphics[width=.9\linewidth]{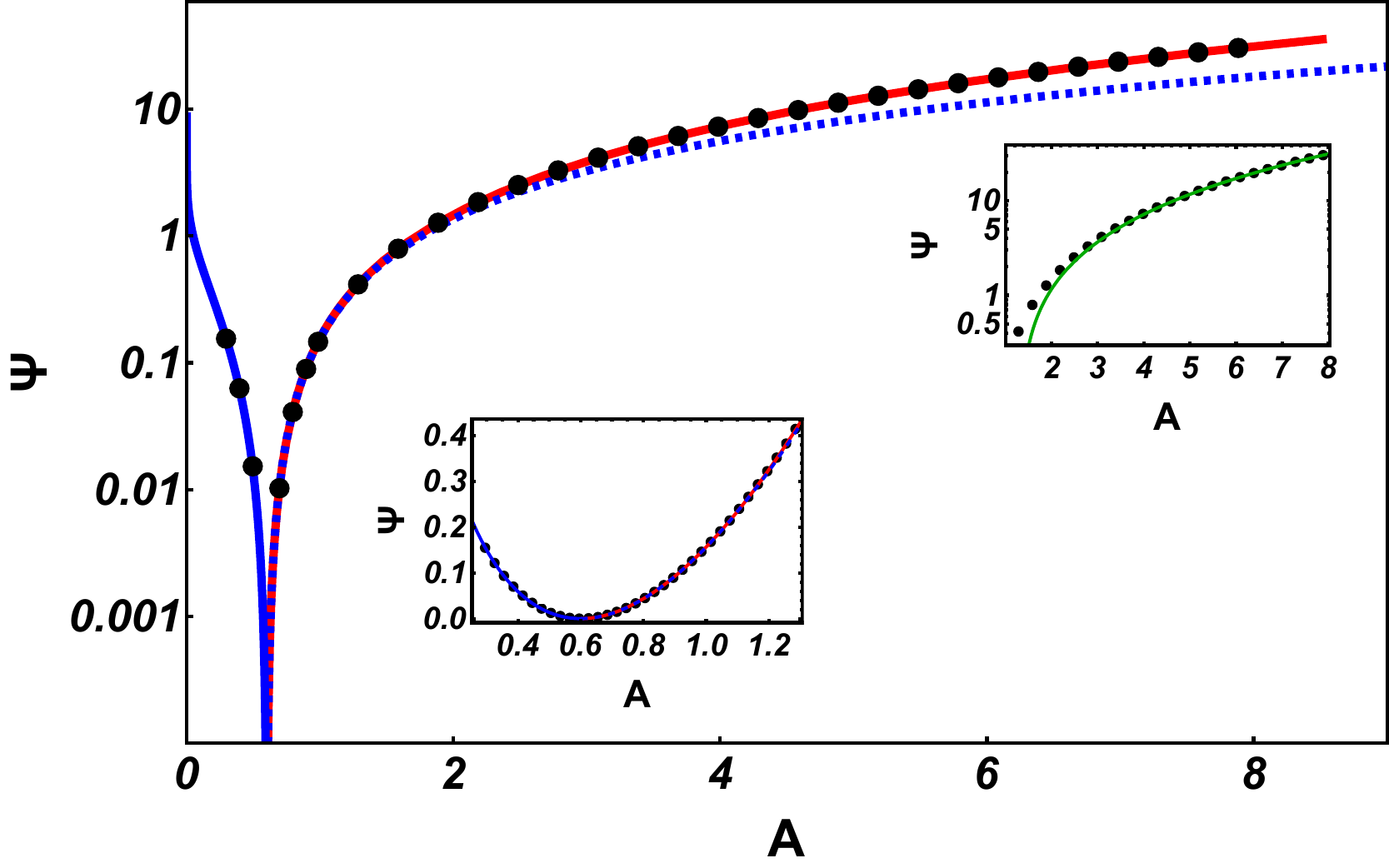}
     \caption{The full rate function $\Psi(A)$ for $m=1$. The left branch (blue) and the right branch (red) correspond to the single- and double-interval solutions, respectively. The black circles show the results of Wang-Landau simulations of the linear statistics of $20\times 20$ GOE matrices. The blue dashed curve shows analytic continuation of a single interval-rate function to the region of $A>\bar{A}$. It deviates from the red curve very slowly, suggesting that the phase transition at $A=\bar{A}$ is of infinite order. The green curve in the right inset is the large $A$ asymptotic (\ref{R_drop}). }
\label{Fig:DI}
   \end{minipage}
\end{figure}

For $m=1$  the integrals in Eq.~(\ref{sol3}) can be evaluated analytically. Noticing that
\begin{equation}
\frac{ \text{sgn} (t)\sqrt{R(t)}(t-\lambda \text{sgn} (t))}{t-x}= \text{sgn} ( t) \sqrt{R(t)}+ \text{sgn} (t) (x-\lambda  \text{sgn} (t))\frac{\sqrt{R(t)}}{t-x},
\end{equation}
and taking into account that $R(-t)=R(t)$, we can rewrite Eq.~\eqref{sol3} as follows:
\begin{equation}
 \frac{1}{B}\pi^2 \sqrt{R(x)}  \text{sgn}(x)\rho(x)=c x+(x-\lambda)\int\limits_{k}^{1}\frac{\sqrt{R(t)}}{t-x}dt-(x+\lambda)\int\limits_{-1}^{-k}\frac{\sqrt{R(t)}}{t-x}dt.
 \label{sol4}
\end{equation}
The integrals
$$
i_+(x,k)=\int\limits_{k}^{1}\frac{\sqrt{R(t)}}{t-x}dt
$$
and
$$
i_-(x,k)=\int\limits_{-1}^{-k}\frac{\sqrt{R(t)}}{t-x}dt
$$
in the r.h.s of Eq.~\eqref{sol4}  can be evaluated analytically, and they are presented, together with the resulting expression for $\mu$, in Appendix \ref{App_A}. Combining Eqs.~\eqref{sol4}, \eqref{i_p} and \eqref{i_m}, we obtain the following  four-parameter solution for $\rho(x;c,\lambda,k,B)$ (for $x\geq 0$):
\begin{equation}\label{rho}
\rho(x;c,\lambda,k,B)=B\frac{cx +(x-\lambda)i_+(x,k)-(x+\lambda)i_-(x,k)}{\pi^2\sqrt{R(x)}}\,.
\end{equation}
The parameters $c,\, \lambda,\, k,$ and $B$ are determined from the normalization condition (\ref{normalization}), the vanishing of the density at the boundaries of support (\ref{BC}), and the condition~(\ref{Im}) on $A$:
\begin{equation}\label{cond}
    \begin{cases}
        \rho(x \to k;c,\lambda,k,B)\to 0\,, \\
        \rho(x \to 1;c,\lambda,k,B)\to 0\,, \\
        \int\limits_{k}^{1} \rho(x;c,\lambda,k,B) dx =1/2\,,\\
        \sqrt{B}\int\limits_{k}^{1} x \rho(x;c,\lambda,k,B) dx =A/2\,.
    \end{cases}
\end{equation}
The first two conditions in Eq.~(\ref{cond}) allow us to eliminate the parameters $c$ and $\lambda$, which yields a two-parameter distribution $\rho(x;k,B)$, some examples of which are shown in Fig. \ref{Fig:Dens}. The last two conditions in Eq.~(\ref{cond}) are hard to implement analytically. Therefore, we calculated the rate function by numerically evaluating the integrals: varying the parameter $k$, determining the parameter $B$ from the normalization condition, and then calculating $A$ and the rate function (\ref{R(A,m)}) for each set of parameters:
\begin{equation}\label{rho_rate}
\Psi_2(B)=\mathcal{E}[\rho(x;k,B)]-\mathcal{E}[\rho_W(x)]=\frac{B}{2}\int\limits_{k}^1 x^2\rho(x;k,B) dx +\frac{B\mu}{2}+\frac{\sqrt{B}\lambda A}{2} -\frac{\ln B}{4}-\frac{1}{8}\left(3+\ln 4\right)\,.
\end{equation}

Figure \ref{Fig:DI} shows the full rate function $\Psi(A)$ for $m=1$. It includes two branches, corresponding to the single-interval solution~\eqref{k_RateB} for $A\leq \bar{A}$ and the double-interval solution  for $A> \bar{A}$, that we have just presented.  Also shown are  results of the Wang-Landau simulations, and excellent agreement is observed.

\subsubsection{Large $A$ asymptotics}

In the general case of $0<m<2$ one can obtain the large-$A$ asymptotic of the rate function without the need for explicit evaluation of the integrals in the r.h.s. of Eq.~(\ref{sol3}). This is because when $k$ approaches 1 (and thus $A\gg\bar{A}$), the double-interval solution $\rho(x;k, B)$ splits into two ``droplets" located far apart, as it is clearly seen from Fig.~\ref{Fig:Dens} for $m=1$. Remarkably, this feature makes it possible to evaluate the rate function's leading term at large $A$ without the need to know the exact form of the solution $\rho(x)$ of Eq.~(\ref{integn}).

Consider a solution $\rho(x)$ of Eq.~(\ref{integn}) which satisfies the normalization condition (\ref{normalization}) and the boundary conditions (\ref{BC}). Let the support of $\rho(x)$ have the form of two droplets at $(-\sqrt{B},-\sqrt{b})\cup(\sqrt{b},\sqrt{B})$, with the distance between them much larger than their sizes, $\sqrt{b}\gg \sqrt{B}-\sqrt{b}$. We can establish simple lower and upper bounds on the linear statistics $A$ of this configuration by replacing $|x|^m$ under the integrals by $b^{m/2}$ and $B^{m/2}= b^{m/2}\left(1+\frac{\sqrt{B}-\sqrt{b}}{\sqrt{b}}\right)^{m}$, respectively:
\begin{multline}
b^{m/2}\left(\;\int\limits_{-\sqrt{B}}^{-\sqrt{b}}+\int\limits_{\sqrt{b}}^{\sqrt{B}}\;\right)\rho(x)dx\leq A=\left(\;\int\limits_{-\sqrt{B}}^{-\sqrt{b}}+\int\limits_{\sqrt{b}}^{\sqrt{B}}\;\right)|x|^m \rho(x)dx\leq \\
\leq b^{m/2}\left(1+\frac{\sqrt{B}-\sqrt{b}}{\sqrt{b}}\right)^{m}\left(\;\int\limits_{-\sqrt{B}}^{-\sqrt{b}}
+\int\limits_{\sqrt{b}}^{\sqrt{B}}\;\right)\rho(x)dx\,.
\end{multline}
Taking into account the strong inequality $\sqrt{b}\gg \sqrt{B}-\sqrt{b}$ and the normalization condition, we obtain the following estimates:
\begin{equation}\label{A_b}
 b^{m/2}\leq A\leq b^{m/2}\left(1+m\frac{\sqrt{B}-\sqrt{b}}{\sqrt{b}}\right)\,.
\end{equation}
In the leading order, we obtain $A\simeq  b^{m/2}$.
Further, we notice that the energy of the two-droplet configuration is dominated by the potential energy $\int x^2 \rho(x) dx \sim b$, while the droplets' interaction energy $\int \ln|x-y| \rho(x) \rho(y) dx dy \sim \ln b$ when the integration involves different droplets. The contribution to the energy of the Coulomb gas coming from the self-interaction term
$\int \ln|x-y| \rho(x) \rho(y) dx dy$, \textit{i.e.}, when the coordinates $x$ and $y$ belong to the same droplet, is of order $\sim{\cal O}(1)$; it does not depend on the distance between the droplets.  The resulting estimate of the rate function's leading term is
\begin{equation}\label{R_drop}
 \Psi_2(A)\simeq\mathcal{E}[\rho(x)]\simeq \int\limits_{\sqrt{b}}^{\sqrt{B}} x^2\rho(x)dx \simeq b\int\limits_{\sqrt{b}}^{\sqrt{B}} \rho(x)dx \simeq  \frac{1}{2}A^{2/m},
\end{equation}
where we have used  the relation $A\simeq  b^{m/2}$, following from Eq.~(\ref{A_b}).

\begin{figure}[H]
     \centering
      \includegraphics[width=.4\linewidth]{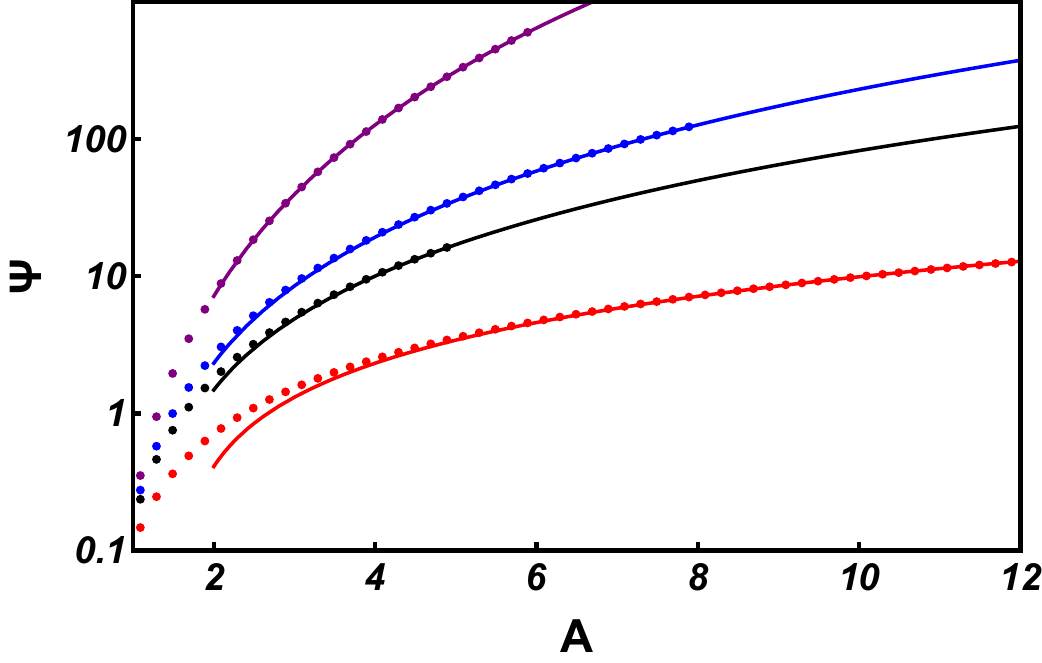}
     \caption{The right tail of the rate function  $N^{-2} \ln \left(\mathcal{P}(A,m)\right)$. The dots represent the results of matrix simulations. The solid lines correspond to Eq.~\eqref{R_drop} for $m=1.5$ (red) $m=0.9$ (black), $m=0.75$ (blue), and $m=0.5$ (purple).}\label{Fig:Data4}
\end{figure}

In Fig. \ref{Fig:Data4} we compare the asymptotic (\ref{R_drop}) with results of the Wang-Landau simulations of the linear statistics of $20 \times 20$ GOE matrices for different values of $m$, and observe a very good agreement.

We reiterate that the asymptotic~(\ref{R_drop}) was obtained without using the solution for the density $\rho(x)$. In the case of $m=1$, however, one can go further and obtain a large-$A$ asymptotic of the optimal density itself, which turns out to be quite instructive.  To this end, it is convenient to consider the force balance equation (\ref{force}):
\begin{equation}\label{DI_solution}
\int\limits_{-\sqrt{B}}^{-\sqrt{b}}\frac{\rho(y)}{x-y}dy+\int\limits_{\sqrt{b}}^{\sqrt{B}}\frac{\rho(y)}{x-y}dy=x-\text{sgn} (x) \Lambda .
\end{equation}
For definiteness, let us consider $x \in \left(\sqrt{b},\sqrt{B}\right)$. Then the first and the second terms in the l.h.s. of Eq.~\eqref{DI_solution} describe the inter-droplet and intra-droplet interactions, respectively. We can establish the lower and upper bounds on the forces $\frac{\rho(y)dy}{x-y}$ acting between droplets  by replacing $x-y$ under the integrals by $2\sqrt{b}$ and $2\sqrt{B}=2\sqrt{b}\left(1+\frac{\sqrt{B}-\sqrt{b}}{\sqrt{b}}\right)$, respectively. We have
\begin{equation}\label{bounds}
\int\limits_{-\sqrt{B}}^{-\sqrt{b}}\frac{\rho(y) dy}{2\sqrt{b}\left(1+\frac{\sqrt{B}-\sqrt{b}}{\sqrt{b}}\right)}\leq \int\limits_{-\sqrt{B}}^{-\sqrt{b}}\frac{\rho(y)dy}{x-y}\leq \int\limits_{-\sqrt{B}}^{-\sqrt{b}}\frac{\rho(y)dy}{2\sqrt{b}},
\end{equation}
or, in the limit  $\sqrt{b}\gg \sqrt{B}-\sqrt{b}$,
\begin{equation}
\frac{1}{4\sqrt{b}}\left(1-\frac{\sqrt{B}-\sqrt{b}}{\sqrt{b}}\right)\leq \int\limits_{-\sqrt{B}}^{-\sqrt{b}}\frac{\rho(y)dy}{x-y}\leq \frac{1}{4\sqrt{b}} \Longrightarrow \int\limits_{-\sqrt{B}}^{-\sqrt{b}}\frac{\rho(y)dy}{x-y}\simeq \frac{1}{4\sqrt{b}},
\end{equation}
That is,  the contribution from the inter-droplet interaction is small compared to that of the intra-droplet interactions. Therefore, the interaction between the droplets is insignificant. This leads us to a simple integral equation
\begin{equation}\label{DI_approx}
\int\limits_{\sqrt{b}}^{\sqrt{B}}\frac{\rho(y)}{x-y}dy\simeq x-\Lambda .
\end{equation}
We immediately notice that the solution of this equation is just a shifted and rescaled semicircular Wigner distribution. Normalizing the total mass to unity, we obtain
\begin{equation}\label{DI_sc}
\begin{cases}
\rho(x,A)=\frac{1}{\pi}\sqrt{(A+1-|x|)(|x|-A+1)}, \\
\sqrt{B}=A+1, \\
\sqrt{b}=A-1, \\
\Lambda = A.
\end{cases}
\end{equation}
In Fig. \ref{Fig:Dens} we compared the full double-interval solution $\rho(x;k, B)$ (\ref{rho}) with the shifted and rescaled Wigner semicircle $(A+1)\rho[x(A +1), A]$ from Eq.~(\ref{DI_sc}). As one can see,  the simple semicircle approximation works surprisingly well even for a relatively small value  $A\simeq 2.98$.

\subsection{Order of the splitting phase transition}
\label{order}

\subsubsection{General}

The change of topology of the support of the optimal density $\rho(x)$ implies a phase transition at $A=A_{\text{cr}}$. In the absence of a closed form
solution for the rate function $\Psi_2(A)$, it is still possible to predict the order of the transition by using some quantitative but non-rigorous argument that we will present shortly. When possible, we will test these predictions numerically.

Our argument makes use of three types of solutions of Eq.~(\ref{inteq2}) for $\rho(x)$ which depend on $m$ and are parametrized by $\Lambda$ and $\mu$. For all the solutions we assume that $\rho(x)$ vanish at the support boundaries. In solutions of one type we will abandon the non-negativity condition. In the solutions of another type we will allow the gas to have mass $M$ different from $1$. These can be obtained from the solutions with $M=1$ via the rescaling transformation ~(\ref{P010}).

The argument starts with the single-interval solutions, $\rho_1(x)$, with $M=1$ but without the non-negativeness condition. They are described by Eqs.~(\ref{1dom}) and~(\ref{AA(B)})-(\ref{mu(B)}), regardless of the sign of $A-A_{\text{cr}}$. According to Eq.~(\ref{rho_exp}), such a solution can be presented in the vicinity of $x=0$ as
\begin{equation}\label{rho_exp2}
\rho_1(x\to 0) {=}
    \rho_0-C(m) \Lambda |x|^{m-1}+{\cal O}(x^2)+{\cal O}(|x|^{m+1}) \qquad \text{for $0<m<2$, but $m\neq 1$}\,.
\end{equation}
Here $C(m)$ is a function of $m$ only, and the sign of $C(m)$ coincides with the sign of $1-m$. The parameter $\rho_0$ depends on $m$, and $\Lambda$. Further arguments depend on whether $0<m<1$ so that $A_{\text{cr}}=\bar{A}$, or $1<m<2$ so that $A_{\text{cr}}>\bar{A}$, see Eq.~(\ref{Acr}).

\subsubsection{$0<m<1$}
\label{mless1}

Here the second  term in Eq.~(\ref{rho_exp2}) is the leading one at $x\to 0$, and it determines the sign of $\rho(x=0)$. Also, here we have $\Lambda_{\text{cr}}=0$, $B_{\text{cr}}=2$, and $A_{\text{cr}}=\bar{A}$. Consider now a solution $\rho_1(x)$ with $0<\Lambda\ll 1$. For this solution the density $\rho_1(x)$ is negative on the interval of the length
\begin{equation}\label{P020}
\Delta \simeq 2\left[\frac{C(m)\Lambda}{\rho_0}\right]^{\frac{1}{1-m}}\sim {\cal O}(1)\, \Lambda^{\frac{1}{1-m}}
\end{equation}
around $x=0$.  This interval gives a negative contribution, $\delta M_1$, to the mass of this solution, $M_1=1$:
\begin{equation}\label{P030}
\delta M_1\sim -\Delta \sim -{\cal O}(1)\, \Lambda^{\frac{1}{1-m}}\, .
\end{equation}
This implies that the contribution of the rest of the gas (where $\rho(x)\geq0$) to the total mass $M_1=1$ is
\begin{equation}\label{P040}
1-\delta M_1 =1 +{\cal O}(1)\, \Lambda^{\frac{1}{1-m}}\, .
\end{equation}
Now consider the solution $\rho_a(x)$ with the same $\mu=\mu_a=\mu_1$ and $\Lambda=\Lambda_1$ as for the solution $\rho_1(x)$, but with the non-negativeness condition restored. As we are keeping the same $\Lambda$ and $\mu$, this solution will have a larger mass $M_a>M_1=1$.  We do not know exact two-interval solutions of Eq.~(\ref{inteq2}) in a closed form. Here we make an assumption that the gap of this solution has a length of the same order of magnitude as in Eq.~(\ref{P020}) and, most importantly, the excess of mass of this solution, $M_2-1$, can be estimated as in Eq.~(\ref{P040}):
\begin{equation}\label{P050}
M_a =1 +{\cal O}(1)\, \Lambda^{\frac{1}{1-m}}\, .
\end{equation}
(The coefficients ${\cal O}(1)$ in Eqs.~(\ref{P040}) and~(\ref{P050}) are quite possibly different, but this does not affect our predictions below.)

Finally, we consider the two-interval solution $\rho_2(x)$ of Eq.~(\ref{inteq2}), which obeys the non-negativeness condition and has the same $\Lambda=\Lambda_2=\Lambda_a=\Lambda_1$. It has, however, a different $\mu=\mu_2\ne\mu_a=\mu_1$ which is tuned so that $M_2=1$.
By virtue of Eq.~(\ref{P050}) it is natural to assume that, in order to bring the total mass $M_2$ to its correct value $1$, one should
change $\mu$ by a quantity ${\cal O}(1)\, \Lambda^{1/(1-m)}$:
\begin{equation}\label{P060}
\mu_2 =\mu_a+{\cal O}(1)\, \Lambda^{\frac{1}{1-m}}\, .
\end{equation}
As a result, the relevant functionals  $A$ and $\Psi$ of the solution $\rho_2(x)$,
get corrected in the same way as $M_2$:
\begin{equation}\label{P070}
A=A_2 =A_1(\Lambda)\left[1+{\cal O}(1)\, \Lambda^{\frac{1}{1-m}}\right]\, ,
\end{equation}
and
\begin{equation}\label{P080}
\Psi=\Psi_2 =\Psi_1(\Lambda)\left[1+{\cal O}(1)\, \Lambda^{\frac{1}{1-m}}\right]\, .
\end{equation}
Importantly, $A_1-\bar{A}$ and $\Psi_1$  are analytic functions of $\Lambda$, which Taylor series start with the terms $\propto \Lambda$ and $\propto \Lambda^2$, respectively. Using these properties and eliminating $\Lambda$ from Eqs.~(\ref{P070}) and~(\ref{P080}), we obtain
\begin{equation}\label{P081}
\Psi (A)_{0<A-\bar{A}\ll1}=\Psi_2(A)=\Psi_1(A)+{\cal O}(1)\, \left(A-\bar{A}\right)^{1+\frac{1}{1-m}}+\dots\,.
\end{equation}
Here we keep only the leading term in $\Psi(A)-\Psi_1(A)$ coming from the leading term of $\Psi_1$ as function of $\Lambda$. The latter one is:
$$\propto \Lambda^2 \sim
\left[A-\bar{A}+{\cal O}(1)\, \left(A-\bar{A}\right)^{\frac{1}{1-m}}\right]^2\sim  \left(A-\bar{A}\right)^2+{\cal O}(1)\,\left(A-\bar{A}\right)\, \left(A-\bar{A}\right)^{\frac{1}{1-m}}.
$$

Since $\Psi_2(A)$ obeys one additional constraint --
non-negativeness -- the second term in Eq.~(\ref{P081}) must be positive.
It immediately follows from Eq.~(\ref{P081}) that the order $p$ of the phase transition at $0<m<1$ is equal to
\begin{equation}\label{P090}
p=1+\frac{1}{1-m}\qquad (0<m<1)\, .
\end{equation}

\subsubsection{$1<m<2$}
\label{mmore1}

In this case the sign of $\rho(x=0)$ for the single-interval solution is determined by the first term, $\rho_0$, in Eq.~(\ref{rho_exp2}). This term vanishes at $\Lambda=\Lambda_{\text{cr}}$, corresponding to $B=B_{\text{cr}}$, see Eq.~(\ref{Acr}).
As before, we start from the single-interval solution $\rho_1(x)$ (again, without the non-negativeness condition) in a small vicinity of $B=B_{\text{cr}}$. According to Eq.~(\ref{rho_exp2}), this solution can be presented in the vicinity of $x=0$ as
\begin{equation}\label{rho_exp3}
\rho_1(x\to 0)=
    \left(\partial_\Lambda\rho_0\right)_{\Lambda=\Lambda_{\text{cr}}}\delta\Lambda-C(m) \Lambda_{\text{cr}} |x|^{m-1}+\dots \qquad \text{for $1<m<2$},
\end{equation}
where $\delta\Lambda=\Lambda-\Lambda_{\text{cr}}$, and both $\left(\partial_\Lambda\rho_0\right)_{\Lambda=\Lambda_{\text{cr}}}$ and $C(m)$ are negative. The parameter $\mu$ of this solution is equal to $\mu_1(\Lambda)$, determined by Eq.~(\ref{mu(B)}). For $\delta\Lambda>0$ the solution $\rho_1(x)$ becomes negative on an interval of length $\Delta$ which can be evaluated, with the help of Eq.~(\ref{rho_exp3}), as
\begin{equation}\label{P100}
\Delta\simeq2\left[\frac{\left(\partial_\Lambda\rho_0\right)_{\Lambda=\Lambda_{\text{cr}}}\delta\Lambda}{C(m)\Lambda_{\text{cr}}}
\right]^{\frac{1}{m-1}}\sim \delta\Lambda^{\frac{1}{m-1}}\, .
\end{equation}
This interval gives a negative contribution, $\delta M_1$, to the total mass $M_1=1$. This contribution can be estimated as
\begin{equation}\label{P110}
\delta M_1\sim \rho_1(0) \Delta\sim -\delta\Lambda^{1+\frac{1}{m-1}}\, .
\end{equation}
This implies that the rest of the gas gives the following contribution to the total mass $M_1=1$:
\begin{equation}\label{P120}
1+{\cal O}(1)\delta\Lambda^{1+\frac{1}{m-1}}\, .
\end{equation}

Now we consider the second solution, $\rho_a(x)$, of Eq.~(\ref{inteq2}) with the same $\Lambda$ and $\mu=\mu_a=\mu_1$, but obeying the non-negativeness condition. This is a two-interval solution with a gap of the size ${\cal O}\Delta$. As previously, our main assumption is that the total mass $M_a$  of this solution can be estimated as in~(\ref{P120}), maybe up to a coefficient ${\cal O}(1)$.

Finally, we consider the third solution, $\rho_2(x)$: a two-interval solution which obeys the non-negativeness condition, which has the same $\Lambda=\Lambda_2=\Lambda_a=\Lambda_1$, but a different $\mu=\mu_2\ne\mu_a=\mu_1$ chosen so as to obey the normalization condition $M_2=1$. Again, it is natural to assume  that, in order to make the total mass equal to $1$, one should
change the parameter $\mu$ by a quantity ${\cal O}(1)\, \Lambda^{1+\frac{1}{m-1}}$:
\begin{equation}\label{P130}
\mu_2 =\mu_a+{\cal O}(1)\, \delta\Lambda^{1+\frac{1}{m-1}}\, .
\end{equation}

\begin{figure}[H]
     \centering
      \includegraphics[width=.3\linewidth]{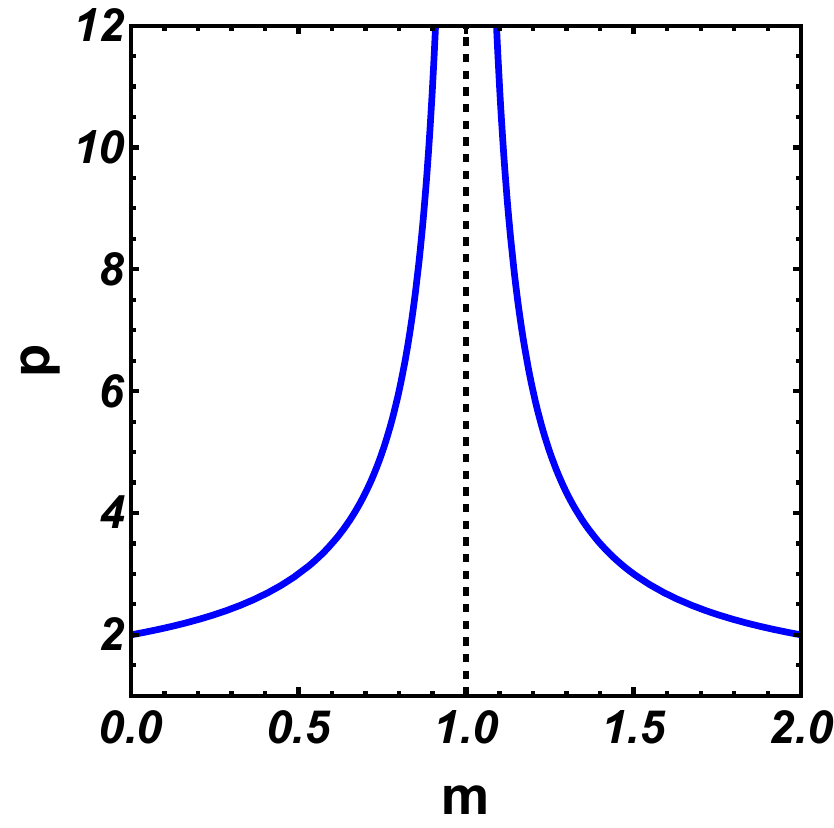}
        \caption{The order of phase transition between the gapless and gapped spectra as a function of $m$.}
        \label{pvsm}
\end{figure}

As a result, the functionals $A$ and $\Psi$ are corrected in the same way as $M$:
\begin{equation}\label{P140}
A=A_2 =A_1(\Lambda)\left[1+{\cal O}(1)\, \delta\Lambda^{1+\frac{1}{m-1}}\right]\, ,
\end{equation}
and
\begin{equation}\label{P150}
\Psi=\Psi_2 =\Psi_1(\Lambda)\left[1+{\cal O}(1)\, \delta\Lambda^{1+\frac{1}{m-1}}\right]\, .
\end{equation}
Now, $A_1(\Lambda)$ and $\Psi_1(\Lambda)$ are analytic functions However, in contrast to the case $0<m<1$, the Taylor series of $A_1(\Lambda)$ and $\Psi_1(\Lambda)$ at the point $\Lambda=\Lambda_{\text{cr}}$ both start from the terms
$A_{\text{cr}}+{\cal O}(1)\, \delta \Lambda$ and $\Psi_{\text{cr}}+{\cal O}(1)\, \delta \Lambda$, respectively. Using these properties and eliminating $\Lambda$ from Eqs.~(\ref{P140}) and~(\ref{P150}), we obtain:
\begin{equation}\label{P160}
\Psi (A)_{0<A-A_{\text{cr}}\ll1}=\Psi_2(A)=\Psi_1(A)+{\cal O}(1)\, \left(A-A_{\text{cr}}\right)^{1+\frac{1}{1-m}}+\dots\, .
\end{equation}
where we have kept only the leading term in $\Psi(A)-\Psi_1(A)$. Again, the second term in Eq.~(\ref{P160}) should be positive. The resulting order $p$ of the phase transition in this case is equal to
\begin{equation}\label{P170}
p=1+\frac{1}{m-1}\qquad \mbox{~~~for~~~}(1<m<2)\, .
\end{equation}
Combining Eqs.~(\ref{P090}) and~(\ref{P170}), we obtain
\begin{equation}\label{P180}
p=1+\frac{1}{|m-1|}\qquad \mbox{~~~for~~~}(0<m<2)\, .
\end{equation}
A plot of the order of this (in general, fractional) phase transition as a function of $m$ is shown in Fig. \ref{pvsm}. In particular, Eq.~(\ref{P180}) indicates that, at $m=1$, the order of the transition is infinite. In this special case, which is formally beyond the validity of our arguments in subsections \ref{mless1} and \ref{mmore1},  the single-interval solution has a logarithmic singularity at $x=0$, see Eq.~(\ref{rho_exp}). Note that similar logarithmic singularities of the optimal density profile have already been associated with infinite-order phase transitions in large deviations of some different (truncated) linear statistics of the Gaussian \cite{Grabsch2022} and Laguerre \cite{Grabsch2017} ensembles of random matrices.

Albeit quantitative, our arguments leading to Eq.~(\ref{P180}) are far from rigorous. Therefore, we tested our predictions numerically. For the special case of $m=1$ we used numerical evaluations of the analytical expressions from Sec. \ref{m1general}. For $m\neq 1$
we employed the Wang-Landau simulations.

\subsubsection{Numerical results: $m=1$}

\begin{figure}[H]
     \centering
      \includegraphics[width=.4\linewidth]{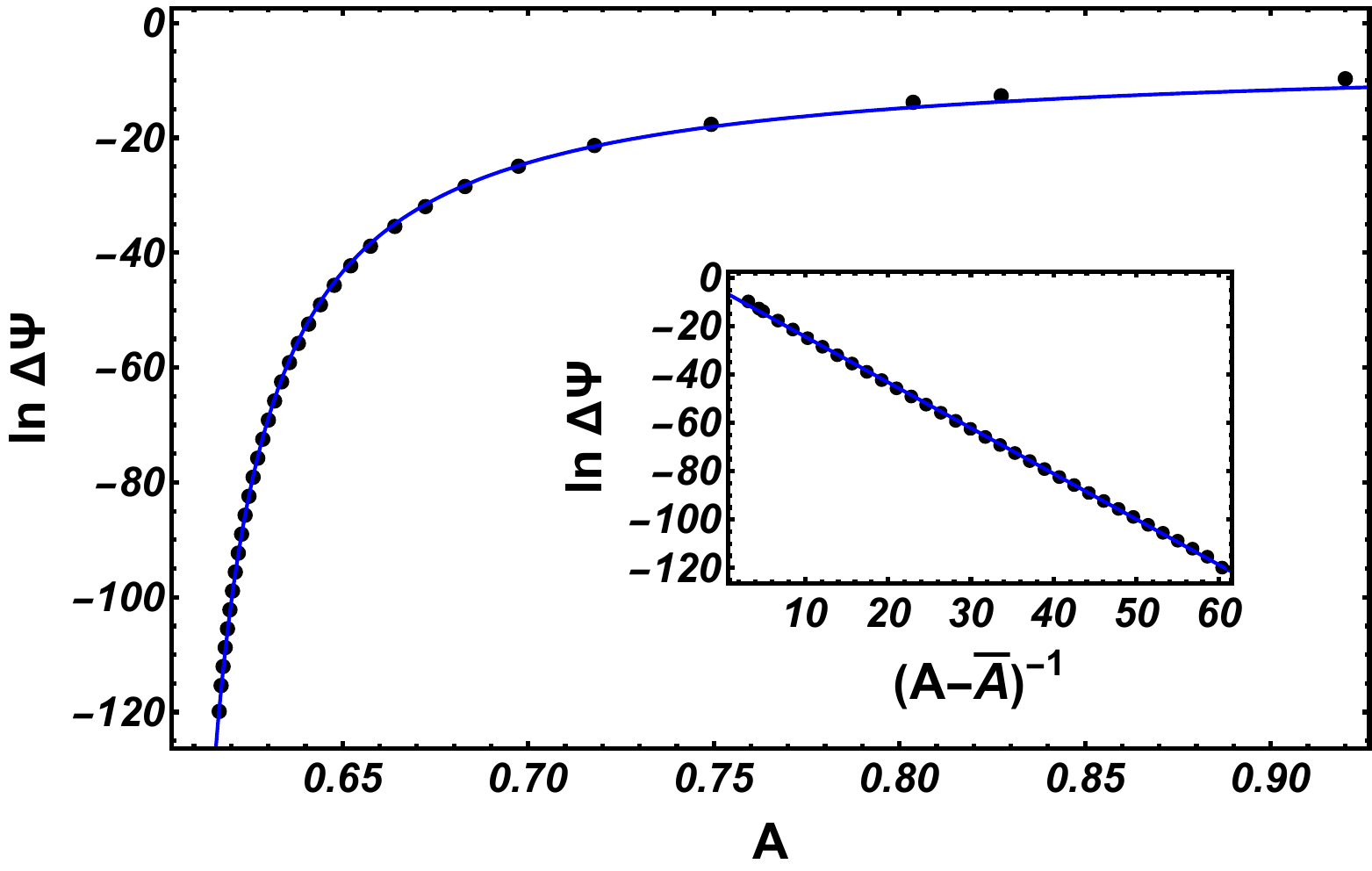}
        \caption{Shown are the difference of the rate functions~(\ref{dR}) (black dots) versus $A$ and its approximation (\ref{expfit}) with $a = 5.40$ and $b= 1.89$ (the blue line). The inset shows the same quantity but as a function of $1/(A-\bar{A})$.}
        \label{Fig:dR}
\end{figure}

To verify our prediction $p(m=1) = \infty$, we inspected the difference between the true rate function (\ref{rho_rate}) for $A>A_{\text{cr}}$, evaluated numerically, and the \emph{analytic continuation} of the single-interval rate function, see Eqs.~(\ref{AA(B)}) and~~(\ref{k_RateB}), to the region of $A>A_{\text{cr}}$:
\begin{equation}\label{dR}
    \Delta \Psi(A)=\Psi_2(A)-\Psi_1(A).
\end{equation}
In Fig.~\ref{Fig:dR} we plot the logarithm of $\Delta \Psi(A)$ as a function of $A$, obtained by numerical evaluation, in the vicinity of $\bar{A}=\frac{4 \sqrt{2}}{3 \pi }= 0.6002\dots$ (black dots). Also plotted is approximation of the form $ \ln \Delta \Psi(A)=-a-b (A-\bar{A})^{-1}$ with the $a = 5.40$ and $b= 1.89$. The observed remarkable agreement strongly suggests that, near the transition point to the right of it, the true rate function of the double-interval solution behaves as
\begin{equation}
\label{expfit}
    \Psi_2(A)\simeq \Psi_1(A) + K\,\exp\left(-\frac{b}{A-\bar{A}}\right)\,.
\end{equation}
with positive $K=e^a$ and $b$.
Noticeable is an essential singularity at the transition point $A=\bar{A}$. As the Taylor series of $\Delta \Psi(A)$ about the transition point vanishes identically, the phase transition  for $m=1$ is of infinite order.

\subsubsection{Numerical results: $0<m<1$}

In order to test out theoretical prediction~(\ref{P081}), we computed numerically, for different values of $m$, the rate function $\Psi(A)=N^{-2}\ln \mathcal{P}(A,m)$ near $A=A_{\text{cr}}$ using the Wang-Landau simulation algorithm and sampling GOE matrices  of size $N=10$. We validated the simulation results by observing a good agreement between them and the results of numerical evaluation of the analytical
formulas for $m=1$, see Fig.~\ref{Fig:DI}.

\begin{figure}[ht]
     \centering
      \includegraphics[width=.7\linewidth]{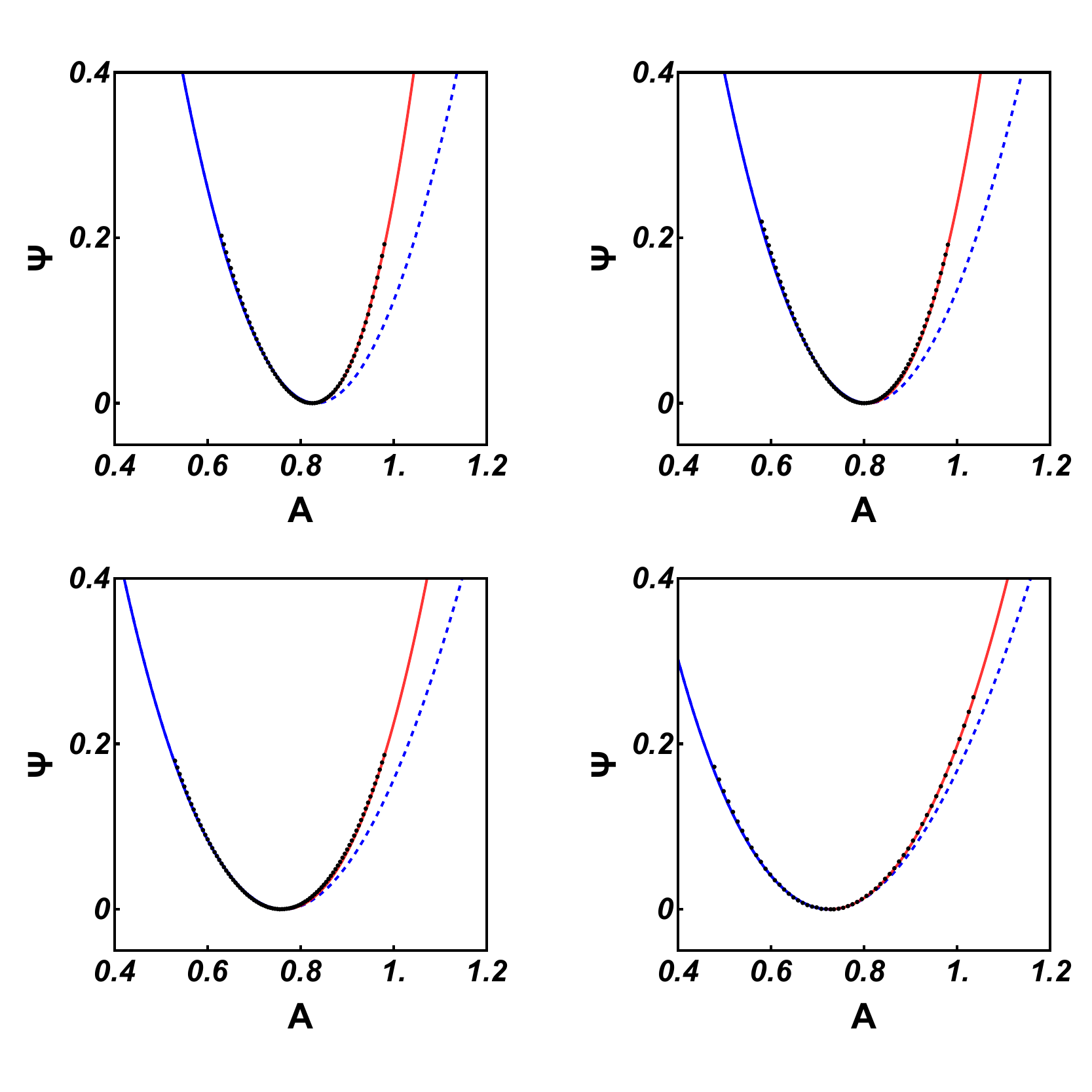}
        \caption{The numerical rate function $N^{-2}\ln \mathcal{P}(A,m)$ (black dots), the single-interval rate function $ \Psi_1$ (blue curve), and the asymptotic (\ref{P081a}) of the double-interval rate function for different values of $m$: (i) Top left: $m=1/4;\, c \simeq 6.86$; (ii) Top right: $m=3/10;\, c \simeq 5.69$; (iii) Bottom left: $m=2/5;\, c \simeq 3.18$; (iv) Bottom right: $m=1/2;\, c \simeq 1.46$.}
        \label{Fig:order}
\end{figure}

According to Eq. (\ref{P081}),  the expected asymptotic behavior of $\Psi(A)$ near the transition point $A_{\text{cr}}$ should be
\begin{equation}
\label{P081a}
    \Psi(A) \simeq \Psi_1(A)+c \,(A-A_{\text{cr}})^{1+\frac{1}{|m-1|}}\,,
\end{equation}
where $c$ is an $m$-dependent constant. Figure \ref{Fig:order} shows the results of matrix simulations for different $0<m<1$. In this case $A_{\text{cr}}=\bar{A}$. The numerically found rate function is shown by black dots. The left branch of the rate function, shown in blue, is the single-interval rate function $ \Psi_1(A)$ (\ref{k_RateB}). The red curves show the asymptotic~(\ref{P081a}) of the double-interval rate function, with the constant $c$ serving as the only adjustable parameter. As one can see, the agreement is very good.

It is also clearly seen on Fig.~\ref{Fig:order} that the higher the order of transition $p=1+1/|m-1|$ is, the closer is the double-interval rate function to the analytic continuation of the single-interval rate function to the region of $A>A_{\text{cr}}$. Consequently, as $m$ approaches $1$,  the comparison between Eq.~(\ref{P081a})  becomes challenging due to inevitable computational errors. For $1<m<2$  numerical verification of Eq.~(\ref{P160}) is also difficult, because in this case the transition point $A_{\text{cr}}$ moves to the distribution tail, see Fig. \ref{Fig:A_mu}. When $m\to 2$,  the transition of order $1+1/|m-1|\to 2$ is expected to occur at $A_{\text{cr}}\rightarrow \infty$, making it inaccessible in numerical simulations

\section{Single-particle scaling of rate function}
\label{evaporation}

As we have seen, for  $m<2$  large values of $A$ are achieved when the optimal charge density of the ensemble splits into two separate ``droplets", and the distance between them grows with $A$.
Such a scenario, however, is impossible for $m>2$ due to a qualitative change in the form of the effective potential $V_{\text{eff}}(x, \Lambda)$, see Eq.~(\ref{inteq2}). Indeed, for $\Lambda>0$ and $m>2$, the double-well form of the effective potential disappears. Instead, the effective potential now has a local minimum at $x=0$, and it is unbounded from below at $|x|\to \infty$.
As a result, for sufficiently large $\Lambda>0$ Eq.~(\ref{inteq2}) does not have physically reasonable solutions.

In this case a different type of optimal solution emerges via the evaporation scenario \cite{KTJ}. In this scenario, one point charge ``evaporates" from the rest of charges and is located at an \textit{a priori} unknown position $x=X$ to be found, while the rest of charges can still be described as a continuous ``sea":
\begin{equation}
    \rho(x)=\tilde{\rho}(x)+\frac{1}{N} \delta(x-X)\,.
\end{equation}
The evaporation changes in the modified energy of the Coulomb gas (\ref{energy1}):
\begin{multline}\label{energy_s}
\mathcal{E}[X,\tilde{\rho}(x),\Lambda,\mu] = \frac{1}{2} \int_{-\infty}^{\infty} x^2\,\tilde{\rho}(x)\,dx
- \frac{1}{2}\int_{-\infty}^{\infty}\int_{-\infty}^{\infty} \ln|x-y| \tilde{\rho}(x) \tilde{\rho}(y) dx\,dy\,\\+\frac{1}{N}\left(\frac{X^2}{2}-\int_{-\infty}^{\infty} \ln|X-x| \tilde{\rho}(x)dx\right)- \Lambda\left(\int_{-\infty}^{\infty} |x|^m \tilde{\rho}(x)\,dx +\frac{1}{N}|X|^m-A\right) - \mu\left(\int_{-\infty}^{\infty}  \tilde{\rho}(x)\,dx+\frac{1}{N}-1\right)\,,
\end{multline}
while the normalization condition (\ref{Im}) and the linear statistics condition (\ref{normalization}) become
\begin{eqnarray}
\int \rho(x) dx &=&\int \tilde{\rho}(x) dx+\frac{1}{N}=1 \label{s_norm}\,,\\
\int |x|^m \rho(x) dx &=&\int |x|^m \tilde{\rho}(x) dx+\frac{1}{N} |X|^m=A \,.\label{s_A}
\end{eqnarray}

The optimal density $\rho(x)$ is a solution of the minimization problem (\ref{Variation}) with an additional condition on the variation of the evaporated particle position $X$:
\begin{equation}\label{A010}
    \frac{\partial \mathcal{E}[X,\tilde{\rho}(x),\Lambda, \mu]}{\partial X}=0\,.
\end{equation}
As in the previous works \cite{Nadal_2011,MajumdarVergassola}, the minimum is achieved when the ``sea"  of particles $\tilde{\rho}(x)$   ``freezes" in the ground state -- in our case in the Wigner semicircle,  whereas the evaporated particle dominates the contribution to $A$ compared with the rest of particles. We have
\begin{equation}\label{rhos}
    \begin{cases}
        \rho(x)=\rho_{\text{W}}(x)+\frac{1}{N} \delta(x-X)\,, \\
        A=\bar{A}+\frac{X^m}{N}=\frac{2^{\frac{m}{2}+1} \Gamma \left(\frac{m+1}{2}\right)}{\sqrt{\pi } (m+2) \Gamma \left(\frac{m}{2}+1\right)}+\frac{X^m}{N}\,,
    \end{cases}
\end{equation}
where we have used Eq.~(\ref{barA}) for $\bar{A}$. Substituting Eqs.~(\ref{rhos}) into Eq.~(\ref{energy_s}), we obtain the total energy of the evaporated charge:
\begin{multline}
\mathcal{E}[X,\rho_{\text{W}}(x)]=\frac{1}{2} \int_{-\sqrt{2}}^{\sqrt{2}} x^2\,\rho_{\text{W}}(x)\,dx
- \frac{1}{2}\int_{-\sqrt{2}}^{\sqrt{2}}\int_{-\sqrt{2}}^{\sqrt{2}} \ln|x-y| \rho_{\text{W}}(x) \rho_{\text{W}}(y) dx\,dy\,\\+\frac{1}{N}\left(\frac{X^2}{2}-\int_{-\sqrt{2}}^{\sqrt{2}} \ln|X-x| \rho_{\text{W}}(x)dx\right)\,.
\end{multline}
Subtracting from here the energy of the most probable configuration -- that is, of the Wigner semicircle -- we obtain the (single-particle) rate function:
\begin{equation}\label{Psi_int}
  \frac{1}{N}\Psi_s(X)=\frac{1}{N}\left(\frac{X^2}{2}-\int_{-\sqrt{2}}^{\sqrt{2}} \ln|X-x|\, \rho_{\text{W}}(x)\, dx +\text{const}\,\right) ,
\end{equation}
where the constant is defined in such a way that $\Psi_s(X=\sqrt{2})=0$ \cite{Majumdar_2014}, indicating that the energy of the particle remaining in the continuous eigenvalue ``sea" is equivalent to that of the Wigner semicircle. Performing the integration in Eq.~(\ref{Psi_int}) and taking into account that $|X|>\sqrt{2}$, we arrive at the following expression for the single-particle rate function $\Psi_s(A)$:
\begin{equation}
\begin{cases} \label{srate}
    \Psi_s(X)=\frac{1}{2}|X| \sqrt{X^2-2}+ \ln \frac{|X|-\sqrt{X^2-2}}{\sqrt{2}}\,, \\
    |X|^m=N(A-\bar{A})\,,
    \end{cases}
\end{equation}
leading to the single-particle rate function of the form
\begin{equation} \label{A020}
-\ln \mathcal{P}(A,m)\simeq \beta N{\Psi_s}\{X=\left[N(A-\bar{A})\right]^{\frac{1}{m}}\}\, .
\end{equation}
In this work we are interested in the limit of $N\to\infty$ with $A-\bar{A}=\mathcal{O}(1)>0$ fixed.
Using the large-$X$ asymptotic of $\Psi_s(X)$ from Eq.~(\ref{srate}), we arrive at the  probability distribution
\begin{equation} \label{A030}
-\ln \mathcal{P}(A,m)\bigr|_{N\to\infty}\simeq\frac{1}{2}\,\beta N^{\frac{2}{m}+1} (A-\bar{A})^{\frac{2}{m}}\,,
\end{equation}
announced in Eq.~(\ref{singlescaling2}). In its turn, the collective scenario predicts, for $N\to\infty$, $A-\bar{A}=\mathcal{O}(1)$ and $m>2$,
\begin{equation} \label{A040}
-\ln \mathcal{P}(A,m)\bigr|_{N\to\infty}=\beta N^2\Psi_1(A)\,,
\end{equation}
see  Eq.~(\ref{k_RateB}). As one can see, at $A-\bar{A}=\mathcal{O}(1)>0$ and $m>2$ the probability~(\ref{A030}) is higher than the probability~(\ref{A040}). Overall, for  $m>2$, $N\to\infty$, and arbitrary $A-\bar{A}=\mathcal{O}(1)$ we obtain
\begin{equation} \label{A050}
-\beta^{-1} \ln \mathcal{P}(A,m) \simeq
\begin{cases}
N^2\Psi_1(A) \qquad \text{for $A<\bar{A}$}\,,\\
\frac{1}{2}N^{\frac{2}{m}+1} (A-\bar{A})^{\frac{2}{m}}\qquad \text{for $A>\bar{A}$}\,,
\end{cases}
\end{equation}
which summarizes our leading-order predictions for  $m>2$. In the limit of $N\rightarrow \infty$ the rescaled rate function $\Psi(A)=\beta^{-1}N^{-2} \ln \mathcal{P}(A,m)$ \emph{vanishes} for $A>\bar{A}$, signaling  a second-order phase transition at the ground state $A=\bar{A}$.

\begin{figure} [ht]
\includegraphics[width=0.8\textwidth,clip=]{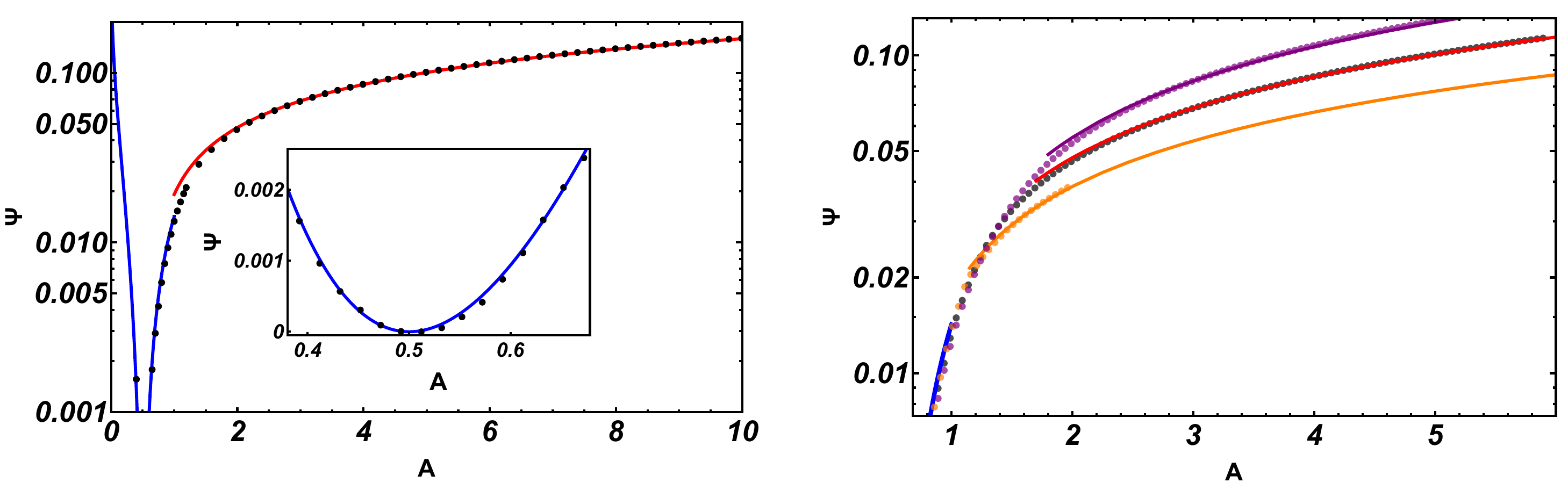}
\caption{Left panel: the simulated rate function $\Psi(A)=N^{-2} \ln \mathcal{P}(A,m)$ (dots) for $m=4$ and $N=60$, the collective rate function $\Psi_1(A)$ (blue), and the single-particle rate function  $N^{-1}\Psi_s(A,N)$, corresponding to the evaporation scenario  (red). The inset shows
a blowup of the vicinity of $A=\bar{A}$ for $m=4$.  Right panel: the right tail of the rate function  $N^{-2} \ln \left(\mathcal{P}(A,m)\right)$. The dots represent the results of GOE matrix simulations for $N=30$ (purple), $N=60$ (black), and $N=120$ (orange). The solid lines represent the single-particle rate functions (\ref{srate}) for the corresponding values of $N$. Note that, for $m=4$, the critical value $A_*$ from Eq.~(\ref{Astar}) is equal to $1$.}
\label{Fig:Data5}
\end{figure}

How do these leading-order predictions compare with the results of our GOE matrix simulations? The latter are shown, for $m=4$ and a moderately large value $N=60$, in the left panel of Fig.~\ref{Fig:Data5}. The dots represent the simulation results for the rescaled distribution $N^{-2} \ln \left[\mathcal{P}(A,m)\right]$. The blue lines show the collective rate function $\Psi_1(A)$ from Eq.~(\ref{k_RateB}), and the red line shows the single-particle rate function $N^{-1}{\Psi_s}\{X=\left[N(A-\bar{A})\right]^{1/m}\}$ from  Eq.~(\ref{srate}). As one can see, the single-particle rate function describes the right tail of the numerical data remarkably well. However, it deviates from the simulated data for $A$ closer to $\bar{A}$. In its turn, the collective rate function $\Psi_1$ continues to be accurate, in a limited range of $A$, for $A>\bar{A}$ where, according to Eq.~(\ref{A050}), it should not apply.

These differences between the asymptotic $N\to \infty$ theory and simulations result from insufficiently large $N$, and they decrease with an increase of $N$. The latter fact is evident on the right panel of Fig.~\ref{Fig:Data5}, which shows the results of matrix simulation for different $N$. As one can see, the range of values of $A$, not covered by either the collective rate function $\Psi_1$, or the single-particle rate function $\Psi_s$, shrinks with an increase of $N$\footnote{\label{collapse}One can notice an additional interesting feature in the right panel of Fig.~\ref{Fig:Data5}. The simulation data for different $N$ appear to collapse into a single curve in a small region of $A>A_*$, where a single-interval collective solution does not exist. The collapse suggests that, for moderately large $N$ like those used in our simulations, the true rate function in this region still exhibits a collective scaling $-\ln \mathcal{P} \sim N^2 f(A)$. In any case, such an intermediate scenario, if it exists, becomes sub-optimal and is replaced by the evaporation scenario as $N$ goes to infinity.}. Unfortunately, it seems impractical to reach the asymptotic regime of Eq.~(\ref{A050}) in the matrix simulations.

\section{$m\gg 1$: from second- to third-order transition}

\label{Lm}

We have just seen that, at $m>2$ the rate function of spectral linear statistics~(\ref{exactsum}) in the limit of $N\to \infty$ exhibits a  phase transition, described by evaporation scenario. This transition occurs at the ground state and, for any finite $m>2$, it is always of second order. A natural question arises on what happens in the limit of $m\to \infty$ when the statistics~(\ref{exactsum}) is dominated by the maximum value of $|\lambda_i|^m$, $i=1,2,\dots, N$.
The statistics of the largest eigenvalue $\lambda_{\max}$ itself (without the absolute value) has been studied in several works, see Ref. \cite{Majumdar_2014} and references therein. For this and other related statistics the evaporation scenario is also at work. In the limit of $N\to\infty$, the evaporation scenario predicts a vanishing largest-eigenvalue rate function $\beta^{-1} N^{-2} \ln \mathcal{P}(\lambda_{\max})$ for $\lambda_{\max}>\sqrt{2}$. However, the associated phase transition, governed by the collective rate function  for $\lambda_{\max}<\sqrt{2}$, is of \emph{third} order, see \textit{e.g.} Ref. \cite{Majumdar_2014}.
To clarify this issue, here we consider in some detail the limit of $m\gg1$  of the single-interval rate function, described by  Eqs.~(\ref{AA(B)}) and~(\ref{k_RateB}).

The limit of $m\to \infty$ can be taken in different ways. Let us introduce $\Delta A \equiv A-\bar{A}$. First we consider the limit of $m\to \infty$ at fixed $\Delta A/\bar{A}=\mathcal{O}(1)$. To this end we use Eqs.~(\ref{barA}) and (\ref{AA(B)}) and present $A/\bar{A}$ in the following form:
\begin{equation}\label{A098}
\frac{A}{\bar{A}}=\left[1+\frac{(m-2)(2-B)}{8}\right]\exp\left[\frac{m}{2}\ln\left(1-\frac{2-B}{2}\right)\right]\, .
\end{equation}
It is clear from Eq.~(\ref{A098}) that, in order to have $\Delta A/\bar{A}=\mathcal{O}(1)$ at $m\to \infty$,
we must demand that $2-B\to 0$ so that $u\equiv m(2-B)=\mathcal{O}(1)$. In the limit of $m\to \infty$
at fixed $u$, we obtain
\begin{equation}\label{A100}
\frac{\Delta A}{\bar{A}}\biggr|_{m\gg 1}=-1 +\frac{u+8}{8}e^{-\frac{u}{4}}+\dots\,, \qquad u>-2\,,
\end{equation}
where we have used the identity
$$
\lim_{m\to \infty}\left(1-\frac{u}{2m}\right)^{\frac{m}{2}} = e^{-\frac{u}{4}}\,.
$$
Under the same scaling $u\equiv m(2-B) =\mathcal{O}(1)$, Eq.~(\ref{k_RateB}) becomes
\begin{equation}\label{A110}
\Psi_1\biggr|_{m\gg 1}
=\frac{1}{m^3}\left(\frac{u^2}{16}+\frac{u^3}{96}\right)+\dots\,.
\end{equation}
Equations (\ref{A100})  and (\ref{A110}) determine in implicit form the $m\gg 1$ asymptotic of the rate function $\Psi_1$ as a function of $\Delta A/\bar{A}$:
\begin{equation}\label{A112}
\Psi_1\left(\frac{\Delta A}{\bar{A}}\right)\biggr|_{m\gg 1}=\frac{1}{m^{3}}f\left(\frac{\Delta A}{\bar{A}}\right)\, .
\end{equation}
The function $f(\dots)$ is defined by Eqs.~(\ref{A100}) and~(\ref{A110}). At small arguments it can be expanded as
\begin{equation}\label{A114}
f\left(\frac{\Delta A}{\bar{A}}\right)=4\left(\frac{\Delta A}{\bar{A}}\right)^2+\dots\, ,
\end{equation}
where we have used the small-$u$ expansion of Eq.~(\ref{A100}),
\begin{equation}\label{A100a}
\frac{\Delta A}{\bar{A}}\biggr|_{u\ll 1}=-\frac{u}{8} + \mathcal{O}(u^3)\,,
\end{equation}
to express $u$ through $\Delta A/\bar{A}$. Equations~(\ref{A112}) and~(\ref{A114}) show that, in the $A$-representation, the phase transition remains of second order even in the limit of $m\to \infty$.

Let us now consider the limit of $m\to \infty$ at constant $2-B=\mathcal{O}(1)$.
Using Eq.~(\ref{k_RateB}), we obtain
\begin{equation}\label{A130}
\Psi_1(B)\bigr|_{m\to\infty}=\frac{1}{4}\ln\frac{2}{B}+\frac{(2-B)(B-6)}{32}\,,\quad (B<2)\, .
\end{equation}
We recall that $B$ is the square of the radius of support of the optimal density distribution for a given $A<\bar{A}$. On the other hand, using  Eq.~(\ref{AA(B)}), we obtain
\begin{equation}\label{A120}
\lim_{m\to\infty}A^{\frac{2}{m}}=B\,,\quad B<2\, .
\end{equation}
That is, in the limit of $m\to \infty$, the statistics of $A^{2/m}$ is equivalent to the statistics of $B$ itself. It is not surprising, therefore, that the rate function (\ref{A130}) can be also obtained, in a straightforward way, from the known rate function \cite{DeanMajumdar_2008} of the joint probability distribution of the minimum and the maximum eigenvalues $\lambda_{\text{min}}$ and $\lambda_{\text{max}}$ of the same class of matrices. One should only set there, for the rescaled eigenvalues, $x_{\text{min}} = - x_{\text{max}} = -\sqrt{B}$, where $0<x_{\text{max}}<\sqrt{2}$.

For small $2-B>0$ Eq.~(\ref{A130}) yields
\begin{equation}\label{A140}
\Psi_1(B, m\to \infty)\bigr|_{0<2-B\ll1}=\frac{(2-B)^3}{96}+ \dots \,.
\end{equation}
As one can see, the transition does become of third order in the limit of $m\to \infty$, but for the order parameter $A^{2/m}$ rather than $A$\footnote{\label{equally}Actually, one can use here $A^{1/m}$ raised to any finite positive power.}.

\

\section{Summary and discussion}
\label{discussion}

As we have shown here, the large-deviation behavior of the probability distribution $\mathcal{P}(A,m)$ at $N\to \infty$ is quite intricate, and the phase diagram of the system on the $(A,m)$ plane, depicted in Fig.~\ref{Fig:phasediagram},  is unexpectedly rich. This is especially true in the region $0<m<2$ of the phase diagram, where we observed a change in topology of the optimal density of the Coulomb gas at $A=A_{\text{cr}}$ accompanied by a phase transition of a collective nature. This phase transition occurs at a critical value of $A$ that we determined. The transition has  an $m$-dependent order $p$, such that $2<p\leq \infty$, see Eq.~(\ref{P180}) and Fig. \ref{pvsm}.

In the future work one can try to improve our non-rigorous analytical arguments leading to the prediction of $p=p(m)$ in the regime of $0<m<2$. An obvious obstacle on the way is
the difficulty in the evaluation, for arbitrary $0<m<2$, of the integrals that appear in the expression (\ref{sol1}) for the exact solution of the two-interval Carleman's equation (\ref{DI solution}).
One can try to circumvent this difficulty by developing a perturbative method of evaluating these integrals which employs from the start
the small parameter $k\ll 1$.

According to our non-rigorous arguments, supported by numerics, the transition order at $0<m<2$ is determined by the type of singularity at $x=0$, exhibited by the single-interval optimal density profile. It is reasonable to assume that a phase transition of the same order  can occur for other choices of linear statistics, confining potential and inter-particle interaction if those lead to the same type of density singularity.  It would be very interesting therefore to look for similar phase-transition phenomena in large deviations of linear spectral statistics in additional systems
which are describable in terms of an ensemble of particles confined by an external potential.  Immediate examples are provided by non-interacting and interacting spinless fermions
in confining potentials \cite{Dean_2019,Smith_2021}, as well as interacting classical gases, not necessarily related to random matrices, such as those studied in Refs. \cite{Agarwal_2019,Kumar_2020}. These examples may include systems where a rigorous determination of the transition order is more straightforward.

For $m>2$, we found a different kind of phase transition, which is described by evaporation scenario. The transition occurs at the ground state of the Coulomb gas and, for any finite $m>2$,  it is always of second order. If one keeps  $A$ as the order parameter, the transition remains of second order even in the limit of $m\to \infty$. However, when switching, \textit{e.g.}, to the order parameter $A^{2/m}$, the transition becomes of third order in this limit.   This happens because, as $m\to\infty$, the order parameter $A^{2/m}$ coincides with $B$, see Eq.~(\ref{A120}). A third-order phase transition of the same nature occurs in the statistics of the maximum (or minimum) eigenvalue,  see Ref. \cite{Majumdar_2014} and references therein.

To conclude, we believe that the unexpected richness of the phase diagram and of the phenomenology of the spectral linear statistics~(\ref{exactsum}), uncovered in this work, justifies
further study of these statistics. In particular, it poses interesting questions about the possibility of similar behaviors in other systems of interacting particles.

\section*{Acknowledgments}

We are very grateful to Eytan Katzav, Satya N. Majumdar and Naftali R. Smith for useful discussions. A.V. and B.M. were supported by the Israel Science Foundation (Grant No. 1499/20).

\bibliography{rsc_JoP}

\appendix
\section{Appendix. $i_+(x,k)$, $i_-(x,k)$ and $\mu$}
\label{App_A}

Evaluation of the integrals $i_+(x,k)=\int\limits_{k}^{1}\frac{\sqrt{R(t)}}{t-x}dt$ and $i_-(x,k)=\int\limits_{-1}^{-k}\frac{\sqrt{R(t)}}{t-x}dt$ in the r.h.s. of Eq.~\eqref{sol4} yield the following expressions:
\begin{multline}\label{i_p}
    i_+(x,k)=\frac{1}{k+1}\bigg(2 K\left(\frac{(k-1)^2}{(k+1)^2}\right) \left(k^2+2 k x-1\right)+(k+1) (k+2 x+1) \Re\left(K\left(\frac{(k+1)^2}{(k-1)^2}\right)\right)+ \\ +4 \left(x^2-k^2\right) \Re\left(\Pi \left(\frac{(k-1) (x+1)}{(k+1) (x-1)},\frac{(k-1)^2}{(k+1)^2}\right)\right)+2 \left(k^2-2 x^2+1\right) \Pi \left(\frac{k-1}{k+1},\frac{(k-1)^2}{(k+1)^2}\right)-(k+1)^2 x E\left(\frac{(k-1)^2}{(k+1)^2}\right)\bigg)
\end{multline}
and
\begin{multline}\label{i_m}
    i_-(x,k)=\frac{1}{k-1}\bigg(2 \left(k^2-2 x^2+1\right) \Pi \left(\frac{k+1}{k-1};\sin ^{-1}\left(\frac{1-k}{k+1}\right),\frac{(k+1)^2}{(k-1)^2}\right) -\\ -4 \left(k^2-x^2\right) \Pi \left(\frac{(k+1) (x-1)}{(k-1) (x+1)};\sin ^{-1}\left(\frac{1-k}{k+1}\right),\frac{(k+1)^2}{(k-1)^2}\right)+(k-1) (k+2 x+1) F\left(\sin ^{-1}\left(\frac{1-k}{k+1}\right),\frac{(k+1)^2}{(k-1)^2}\right)+ \\+(k-1)^2 x E\left(\sin ^{-1}\left(\frac{1-k}{k+1}\right),\frac{(k+1)^2}{(k-1)^2}\right)\bigg),
\end{multline}
where $\Pi(n,m)$, $F(z,m),\, \Pi(n;z,m)$ and $E(z,m)$ are complete and incomplete elliptic integrals.

The relation
\begin{equation}
    c= -
\frac{1}{\ln(2/k')} \int\limits_{-1}^{-k}+\int\limits_{k}^{1} \frac{t \,\text{sgn}(t) f(t) dt}{\sqrt{R(t)}}\,,
\end{equation}
which we introduced to simplify the calculation of the optimal density, uniquely determines $\mu$. In the case of $m=1$ one obtains the following expression for $\mu$:
\begin{equation}
\mu=\frac{1}{4 \pi }\left(4 c \ln \left[\frac{2}{\sqrt{1-k^2}}\right]-8 \lambda \Im \left[ E\left(\arcsin \left[\frac{1}{k}\right],k^2\right)-  F\left(\arcsin \left[\frac{1}{k}\right],k^2\right)\right]+\pi  k^2+\pi \right),
\end{equation}
where $\Im\left[\ldots\right]$ denotes the imaginary part of the corresponding complex function.

\end{document}